\documentclass[preprint, 10pt]{aastex}
 
\begin{document}

\title{Morphological Classification of Galaxies by Shapelet Decomposition in the Sloan Digital Sky Survey II: Multiwavelength Classification}
\author{Brandon C. Kelly}
\affil{Steward Obvservatory, Tucson, AZ 85721-0065 \\ and \\ University of Michigan, Ann Arbor, MI 48109-1090}
\email{bkelly@as.arizona.edu}
\author{Timothy A. McKay}
\affil{Physics Department, University of Michigan, Ann Arbor, MI 48109-1090}
\email{tamckay@umich.edu}

\begin{abstract}

We describe the application of the `shapelet' linear decomposition of galaxy images to multi-wavelength morphological classification using the $u,g,r,i,$ and $z$-band images of 1519 galaxies from the Sloan Digital Sky Survey.  This combination of morphological information in a variety of bands is unique, and it allows automatic separation of different classes in ways which is impossible using single band images or simple spectro-photometric measurements such as color. We utilize elliptical shapelets to remove to first-order the effect of inclination on morphology.  After decomposing the galaxies we perform a principal component analysis on the shapelet coefficients to reduce the dimensionality of the spectral morphological parameter space.  We give a description of each of the first ten principal component's contribution to a galaxy's spectral morphology.  We find that galaxies of different broad Hubble type separate cleanly in the principal component space. We apply a mixture of Gaussians model to the 2-dimensional space spanned by the first two principal components and use the results as a basis for classification.  Using the mixture model, we separate galaxies into three classes and give a description of each class's physical and morphological properties.  Galaxies were typically robustly classified, with $80\%$ of galaxies having a probability of $\geq 90\%$ of occupying their respective class. We find that the two dominant mixture model classes correspond to early and late type galaxies, respectively, both in their morphology and their physical parameters (e.g., color, velocity dispersions, etc.). The third class has, on average, a blue, extended core surrounded by a faint red halo, and typically exhibits some asymmetry.  The third class cannot be associated with any broad Hubble type, however it is the most probable class for irregular galaxies. We compare our method to a simple cut on $u-r$ color and find the shapelet method to be superior in separating galaxies. Furthermore, we find evidence that the $u-r=2.22$ decision boundary may not be optimal for separation between early and late type galaxies, and suggest that the optimal cut may be $u-r \sim 2.4$. We conclude with a discussion of the limitations of our method and ways in which it may be improved. Our framework provides an objective and quantitative alternative to traditional one color visual classification, and the powerful use of both spectral and morphological information gives our method an advantage over separation techniques based on simpler calculations.

\end{abstract}


\section{INTRODUCTION}

\label{intro}

Morphological classification has remained an active and fundamental area of extragalactic astronomy.  Traditional classification based on the Hubble sequence \citep{hub} has played an important role in the development of the morphological study of galaxies.  As data sets grow larger and more precise, the Hubble scheme is becoming increasingly inadequate as a framework in which to do morphological classification \citep[e.g, see][]{cons, bergh01, abr96a}.  Hubble based his system on the $B$-band morphologies of galaxies, and classification has almost exclusively relied on $B$-band data.  Early multi-wavelength studies of galaxies include the discovery of ``red arms'' by \citet{zwicky} and a comparison of the disk and arm structure of six spirals by \citet{schwz}.  In recent years, numerous studies have been done comparing the optical and near-infrared (near-IR) morphologies of galaxies.  \citet{block91} found that the optical morphology of NGC309 is that of a multi-arm spiral, whereas its 2.1 micron morphology is that of a two-arm spiral with a prominent central bar. \citet{colb} searched optical and near-IR images of a sample of isolated and group early-type galaxies for shells and other morphological features that provide clues to galaxy evolution. \citet{esk} have shown that, on average, galaxies with $B$-band classifications Sa through Scd appear about one ``T-type'' \citep{deVauc} earlier in the $H$-band, albeit with large scatter. \citet{jarr} finds that galaxies appear smaller in the near-infrared as compared to the optical, and have higher $K$-band surface brightness than $B$-band. In particular, \citet{jarr} concludes that early types appear redder than late types when comparing the $K$- and $B$-band surface brightness.

Because observations in different bands probe different stellar populations, there is significant motivation for developing a multi-wavelength morphological classification system.  It is well known that for the case of spiral galaxies, the shorter wavelength (e.g., the $B$-band) morphology is dominated by knotty regions of young stars and star-forming regions, while the longer wavelength morphology is dominated by older stars with a smoother spatial distribution.  For example, \citet{whyte} find that late type galaxies are more asymmetric in the $B$-band than in the $H$-band, representing the prevalence of the patchy, irregular star forming regions at shorter wavelengths.  In addition, they find that galaxies are more concentrated in the IR than in the optical; however, this may be either the result of a difference in optical depth between the two bands near the center of galaxies, or because there is a strong color gradient.  Morphological classification based on images in just one observing band is unable to take advantage of variations in the stellar population of the galaxy.  Furthermore, differences in absorption between observing bands can result in structures being obscured by dust, sometimes leading to noticeably different morphologies \citep{esk, block99}.

The inherent subjectivity in the Hubble framework has motivated the development of new and quantitative morphological classification schemes.  The most common of these is based on a central concentration and asymmetry measurement \citep{morgan, doi, abr96b, cons}.  Recent work has investigated utilizing the Gini coefficient for quantifying morphology \citep{abr03,lotz}.  With the advent of large astronomical databases, such as the Sloan Digital Sky Survey \citep[SDSS, ][]{york00}, manual classification is becoming extraordinarily impractical.  Neural networks have proven to be effective at replicating the Hubble classifications of human classifiers \citep{ode, ball}, although in and of themselves they are unable to create a new and quantitative morphological classification framework.  In addition, because of their `black box' nature the results of neural networks are difficult to interpret. Furthermore, many of the proposed classification schemes have only been applied to data from one observing band.  A multi-wavelength quantitative classification of spiral bars was developed by \citet{whyte}, and both \citet{whyte} and \citet{laug} have investigated the dependence of central concentration and asymmetry measurements with wavelength.  In addition, \citet{abr03} and \citet{lotz} have described the dependence of their methods on observing band.  A near-IR ($2.1 \mu$m) classification scheme was developed by \cite{block99} based on the pitch angle and Fourier modes of a spiral galaxy.  Their classification scheme was developed for the purpose of developing a dust-penetrating morphological classification system with emphasis on the the Population II stars, and they did not find any relationship between their near-IR classes and the optical Hubble classes.  While useful, non-optical classification frameworks are still based on one observing band and are incomplete in the sense that it facilitates different classifications for different parts of the spectrum; i.e., independent classification schemes are developed for the optical and near-IR morphologies.

In our previous paper \citep[][hereafter Paper I]{kelly}, we set forth a new classification scheme based on the shapelet \citep{ref} decomposition of the $r$-band images of a volume-limited sample of $\sim 3000$ SDSS galaxies.  We applied a Karhunen-Loeve (KL) transform (or principal component analysis, PCA) to the shapelet coefficients and used a mixture of Gaussians model to estimate the density of the galaxies in the space spanned by the first nine KL modes.  The mixture model was used as a classification framework, where each Gaussian is identified with a morphological class.  Developing a morphological classification system in this manner has the advantage of being model-independent, quantitative, and automatic.  Motivated by the results of our previous analysis, and by the advantages of a multi-wavelength classification scheme, we have performed a similar analysis of SDSS galaxies using the images from all five bands: $u, g, r, i, $ and $z$.  In addition, because we used circular shapelets in the decompositions of Paper I, the axis ratio information was found to contaminate most, if not all, of the principal components.  This is obviously undesirable, as axis ratio is most strongly a result of a galaxy's orientation along our line of sight, and not of its intrinsic morphology. To remedy this, we perform our analysis in this paper with elliptical shapelets, removing to first order the effect of inclination on morphology.

The outline of this paper is as follows. In \S~\ref{shapelets} we give a brief description of the shapelet basis. In \S~\ref{data} we describe our samples and in \S~\ref{decomp} we describe our shapelet decomposition method. In \S~\ref{pca} we describe principal component analysis, as well as show and describe the first ten spectro-morphological principal components. In \S~\ref{the_method} we describe using mixture models for classification, describe each of the three mixture classes, and compare the results with $u-r$ classification. In \S~\ref{conclusions} we conclude with a summary of our results and a discussion describing the limitations of our technique and ways in which it can be improved upon.

\section{SHAPELETS}

\label{shapelets}

Most of the shapelet formalism can be found in \citet{ref}, and Paper I describes the necessary information for our work.  For completeness, we summarize a few of the important points.

Shapelets form a complete orthonormal set, and happen to be the eigenstates of the quantum harmonic oscillator Hamiltonian.  The 1-dimensional basis functions are
\begin{equation}
B_n (x ; \gamma) \equiv \gamma^{-1/2} \phi_n (x / \gamma), 
\label{eq01}
\end{equation}
where $\gamma$ is a characteristic scale, $n$ is a non-negative integer denoting the order, and the dimensionless basis functions $\phi_n$ are
\begin{equation}
\phi_n (x) \equiv \left [ 2^n \pi^{1/2} n! \right]^{1/2} H_n (x) e^{-x^2 /2}.
\label{eq02}
\end{equation}
Here, $H_n (x)$ is a Hermite polynomial of order $n$.  Shapelets are orthonormal over $(-\infty, \infty)$.

\begin{figure}
\begin{center}
\scalebox{0.7}{\rotatebox{90}{\plotone{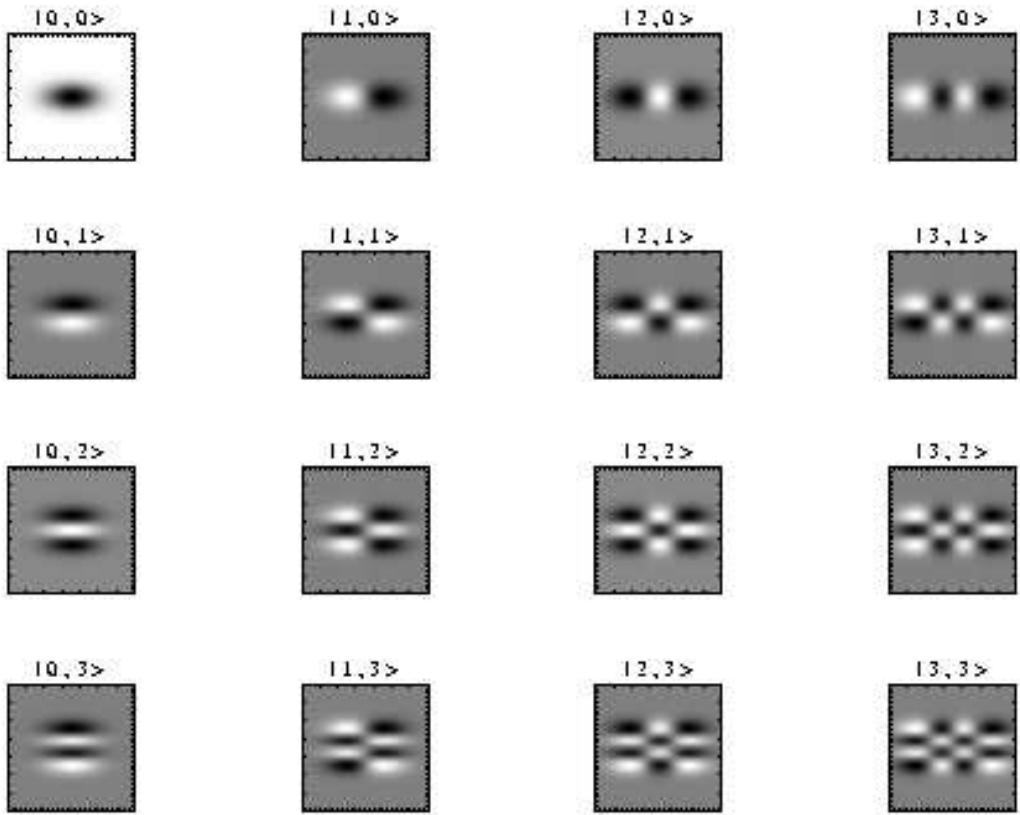}}}
\caption{The first several elliptical shapelets. In all of the images in this paper dark areas correspond to higher values. \label{shapelets_fig}}
\end{center}
\end{figure}

The 2-dimensional shapelets are easily constructed from the 1-dimensional:
\begin{equation}
B_{n_1, n_2} (x_1 / \gamma_1, x_2 / \gamma_2) \equiv (\gamma_1 \gamma_2)^{-1/2} \phi_{n_1} (x_1 / \gamma_1) \phi_{n_2} (x_2 / \gamma_2).
\label{eq03}
\end{equation}
Any sufficiently well behaved function (e.g., a galaxy image) can be decomposed into a sum of shapelets as
\begin{equation}
f ({\bf x}) = \sum_{n_1, n_2 = 0}^{\infty} f_{\bf n} B_{\bf n} ({\bf x}; {\bf \gamma}),
\label{eq04}
\end{equation}
with shapelet coefficients found from the orthonormality property:
\begin{equation}
f_{\bf n} = \int_{-\infty}^{\infty} f ({\bf x}) B_{\bf n} ({\bf x}; {\bf \gamma}) d^2 x.
\label{eq05}
\end{equation}
Here, we have used the notation ${\bf n} = (n_1, n_2), {\bf x} = (x_1, x_2),$ and ${\bf \gamma} = (\gamma_1, \gamma_2)$  In addition, we will also use Dirac notation to denote the shapelet states, where the $n^{th}$ state is denoted as $|n \rangle$ and has x-space representation $\langle x | n \rangle = \phi_n (x)$.  Figure \ref{shapelets_fig} shows the first several elliptical shapelets.

\section{THE DATA}

\label{data}

The SDSS \citep{york00} is an imaging and spectroscopic survey of the Northern Galactic Cap over $\pi$ steradians.  A 2.5m telescope at the Apache Point Observatory, Sunspot, New Mexico, observes the sky in five bands \citep[{\it u, g, r, i, z},][]{fuk96,hogg01,smith02} between 3000 and 10000 \AA , using a drift-scanning mosaic CCD camera \citep{gunn98}, which detects objects to a flux limit of $r \sim 22.5$ mags.  The survey, when finished, is expected to spectroscopically observe 900,000 galaxies down to $r_{lim} \approx 17.77$ mags \citep{str02}, 100,000 Luminous Red Galaxies \citep{eis01}, and 100,000 quasars \citep{ric02}.  The spectroscopic follow-up uses two digital spectrographs on the same telescope as the imaging camera, and the spectroscopic samples are assigned plates and fibers using an algorithm described by \citet{blan03t}.  The astrometric calibration is described in \citet{pier03}.  Details of the galaxy survey can be found in the galaxy target selection paper \citep{str02}, and other principles of the survey are described in the Early Data Release \citep[EDR, ]{sto02}.  Details of the First Data Relase (DR1) can be found in \citet{aba03}

As in Paper I, we use two samples in this analysis.  We first investigate the shapelet method using the $u, g, r, i, $ and $z$-band data for 184 of the 1482 well-resolved galaxies used by \citet[][hereafter Sample 1]{nak03} to estimate the morphology-dependent luminosity function.  We have chosen this group in order to compare the shapelet results with traditional Hubble type, as this catalog contains manual classifications of Hubble type.  The manual classifications make no distinction between spirals with bar structure and spirals without.  Sample 1 contains those galaxies of the \citet{nak03} sample that have redshifts $z < 0.07$ and a PSF $FWHM$ of less than 2.0 kpc when projected onto the plane of the galaxy in all five bands.  This allows us to smooth all galaxies of Sample 1 to a constant scale before decomposing them (see \S~\ref{decomp}).  These galaxies are included in the SDSS EDR.

\begin{figure}
\begin{center}
\scalebox{0.7}{\rotatebox{90}{\plotone{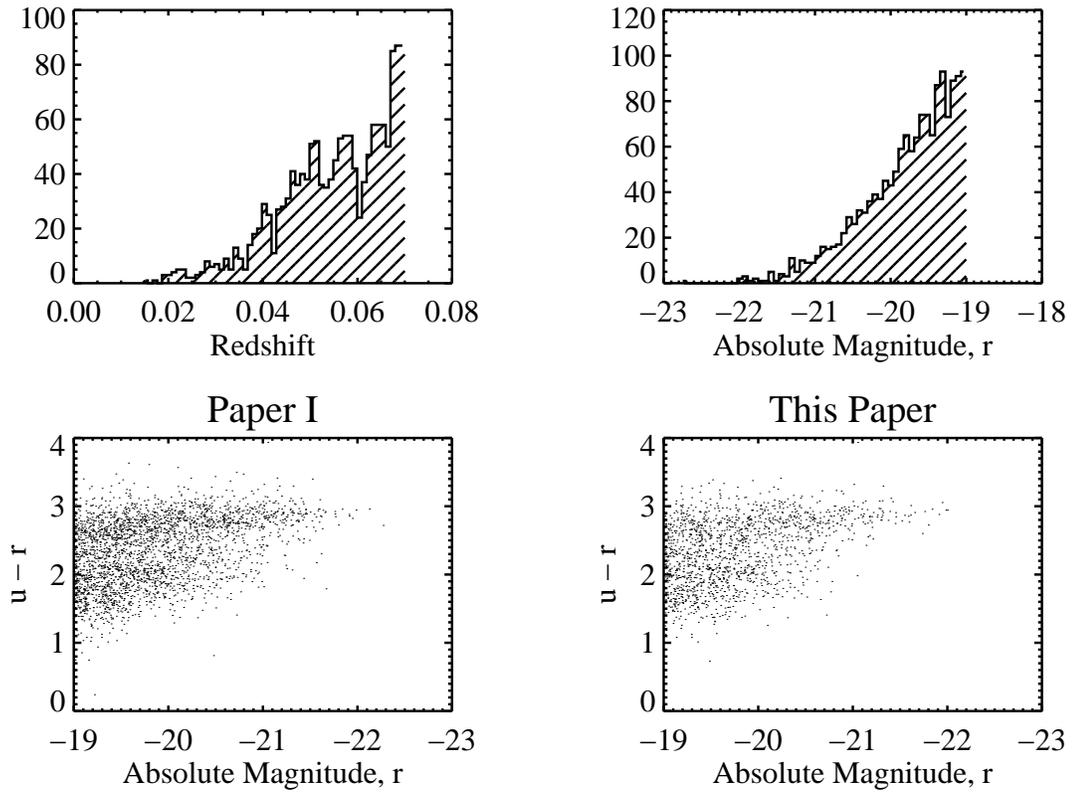}}}
\caption{The distribution of redshift, $r$-band absolute magnitude, and $u - r$ color for the galaxies of Sample 2. Note that there are no large systematic differences between the sample of galaxies used in Paper I and that used in this analysis. \label{sample}}
\end{center}
\end{figure}

We next investigate the shapelet method on the images from all five bands of a volume-limited sample of 1519 nearby galaxies (hereafter Sample 2) included in the SDSS DR1.  These galaxies were chosen because they have projected PSF widths of less than our desired resolution of 2.0 kpc, allowing us to smooth them to this scale in all five bands.  They have redshifts $z<0.07$ and absolute magnitudes $M_u \leq -16, M_g \leq -18$ and $M_r, M_i, M_z \leq -19$.  We made the redshift cut at $z=0.07$ because projected PSF widths become larger than 2.0 kpc for redshifts greater than this, and we chose galaxies of these absolute magnitudes to allow a uniform distribution of absolute magnitude with redshift.  Although we started with the same galaxies as in Paper I, we were left with only $\sim 1600$ galaxies after making these same cuts on all five bands, as opposed to $\sim 3000$ after making the cuts on only the $r$-band data.  In addition, we omitted galaxies for which information necessary to compute the shapelet coefficients (e.g., the SDSS ellipticity parameters) was missing or had large errors in their shapelet reconstruction in at least one of the five bands, leaving a total of 1519 galaxies.  Figure \ref{sample} displays the redshift, $r$-band absolute magnitude, and $u - r$ color distributions for this sample.

\section{DECOMPOSITION METHOD}

\label{decomp}

Most of the information regarding our decomposition method can be found in Paper I; here we reiterate the important points as well as the additions and modifications to the method we have introduced.  We calculate the position angle of the galaxy, $\theta_{pos}$, from the $r$-band SDSS ellipticity parameters, $e_1$ and $e_2$, and its axis ratio, $b/a$, from $e_1$ and $e_2$ in each band. The ellipticity parameters are calculated from the galaxy's adaptively weighted moments \citep{fisc00,bernst} :
\begin{eqnarray}
\theta_{pos} & = & \frac{1}{2} \arctan \left ( \frac{e_2}{e_1} \right ) \nonumber \\
b / a & = & \left ( \frac{1 - e_1 / \cos(2 \theta_{pos})}{1 + e_1 / \cos(2 \theta_{pos})} \right ).
\label{eq06}
\end{eqnarray}
We rotate the image for each band by $\theta_{pos}$, as well as that of the SDSS PSF.  This assures that all galaxies are oriented with their $r$-band major axis oriented along the horizontal.

In Paper I, we used shapelets characterized by a single scale, $\gamma$.  While the use of circular shapelets includes the necessary morphological information for classification, information regarding the galaxy's axis ratio contaminates the principal components and the resulting classification scheme.  To remedy this problem, we use shapelets of varying ellipticity, where the scales along the major and minor axis are $\gamma_1$ and $\gamma_2$, respectively.  The axis ratio of the shapelets is then $b/a = \gamma_2 / \gamma_1$.  For each band, the shapelet scales, $\gamma_1$ and $\gamma_2$, are calculated as
\begin{eqnarray}
\gamma_1 & = & \left [ \frac{I_{xx} + I_{yy}}{1 + (b/a)^2} \right ]^{1/2} \nonumber \\
\gamma_2 & = & \left ( \frac{b}{a} \right ) \gamma_1,
\label{eq07}
\end{eqnarray}
where $I_{xx}, I_{yy}$ are the adaptively weighted moments in the horizontal and vertical directions in that band, and $b/a$ is the axis ratio of the galaxy in that band.  We also compute the PSF axis ratio and scales, $\beta_1$ and $\beta_2$, from the PSF adaptive moments in the same manner as for the galaxy. If $\gamma_i < \beta_i$, then $\gamma_i$ is set to $\beta_i$.  If necessary, we pad the image with blank sky out to a distance $12.5\gamma_i$ from the center, and use the value of $\sigma_{sky}$ for that band to add artificial noise.  This ensures the orthogonality of the shapelets.  We subtract off the sky prior to decomposition.

As outlined in Paper I, we desire to have all galaxies resolved on the same physical scale, ensuring that differences in shapelet coefficients are not the result of differences in resolution.  To do this, we artificially redshift the galaxies to $z = 0.07$, which defines the upper limit of our sample.  Artificial redshifting is performed by rebinning the image such that the angular size of the image, and thus the number of pixels it occupies, is reduced to what it would be if the galaxy were observed at $z=0.07$.  Before artificially redshifting the galaxies, we deconvolve them with the SDSS PSF. This is accomplished by first calculating the shapelet coefficients for the galaxy, $h_{\bf n}$, and the SDSS PSF, $g_{\bf n}$, using the scales given above in each band and decomposing about their respective centers up to a maximum order $n_1 + n_2 = n_{max}$.  The deconvolved shapelet coefficients, $f_{\bf n}$, are then given by
\begin{equation}
f_i = \sum_{j} G^{-1}_{i, j} h_j .
\label{eq08}
\end{equation}
Here, the sum is over the dummy index $j$, and $G^{-1}_{i,j}$ is the inverse of the 'PSF Matrix', $G_{i,j}$, which may be calculated from the shapelet convolution tensor and PSF shapelet coefficients, $g_{\bf n}$ as outlined in \citet{refbac}. The dummy induces $i$ and $j$ are introduced to enable the matrix multiplication by coding the shapelet orders as ${\bf n}(i) = (n_1 (i), n_2(i))$, i.e., $f_i = f_{n_1(i), n_2(i)}$.  The deconvolved image is then constructed from the deconvolved shapelet coefficients, $f_{\bf n}$, with scale $(\gamma_1, \gamma_2)$, as used for the original image. \citet{refbac} found that using the same scale for the deconvolved image as for the original image gave the best results. We then artificially redshift the deconvolved image to $z=0.07$. It should be noted that the deconvolution will appear poor in the pixel-space representation of the galaxy.  This is because we only decompose the galaxy image up to a maximum order of $n_{max}$, and thus are obtaining information from the galaxy between scales of $\theta_{min} \sim \gamma (n_{max} + 1)^{-1/2}$ and $\theta_{max} \sim \gamma (n_{max} + 1)^{1/2}$. Because we use an undercomplete shapelet basis, information outside of these scales is only partially contained within the coefficients. To be more precise, consider some feature in the galaxy image of size less than $\theta_min$.  One is not able to accurately fit this feature because the shapelets are not of high enough order. While information regarding this feature will exist partially in the shapelet coefficients, when one attempts to reconstruct this feature from these coefficients it will appear broadened. This is not a problem for our purposes, as we decompose the smoothed, artificially redshifted galaxy using the same $n_{max}$, and thus the small-scale features are already broadened. Structures on scales smaller than $\theta_{min}$ are unlikely to contribute significantly to spectro-morphological classification as these structures are more likely a product of a galaxy's unique history, rather than the result of physical processes common to all galaxies of a spectro-morphological class.  In this analysis we use $n_{max} = 15$, and thus information of structure less then $\theta_{min} \sim \gamma / 4$ is not completely included in the classification scheme. This corresponds to the shapelet coefficients containing information between $\sim$0.25--4 kpc for the smallest galaxies in our sample and $\sim$2--32 kpc for the largest. We have experimented with using higher values of $n_{max}$, but did not see any noticeable differences in our results that justified the significantly higher computational time.

We convolve the redshifted galaxies in each band with a Gaussian of standard deviation $\beta_0$.  The new shapelet scales, $\gamma'_i$, become $\gamma'_i = \sqrt{\hat{\gamma}^2_i + \beta_0^2}$, where $\hat{\gamma}_i$ is the shapelet scale after reducing $\gamma_i$ to account for the artificial redshifting. We use $\gamma'_i$ to correct for the loss of resolution resulting from the additional smoothing, ensuring that $\gamma'_i > \beta_0$.  It should be noted that although errors are introduced from the deconvolution, most notably in the higher order coefficients, the additional convolution by a Gaussian `smooths out' these errors, as the effect of the convolution process is to project the higher order coefficients onto the lower order ones \citep{ref}. We use a standard physical width in resolution of $\beta_0 = 0.8493$ kpc, corresponding to a Gaussian PSF $FWHM$ of $2.0$ kpc as projected onto the plane of the galaxy.

\begin{figure}
\begin{center}
\scalebox{0.7}{\rotatebox{90}{\plotone{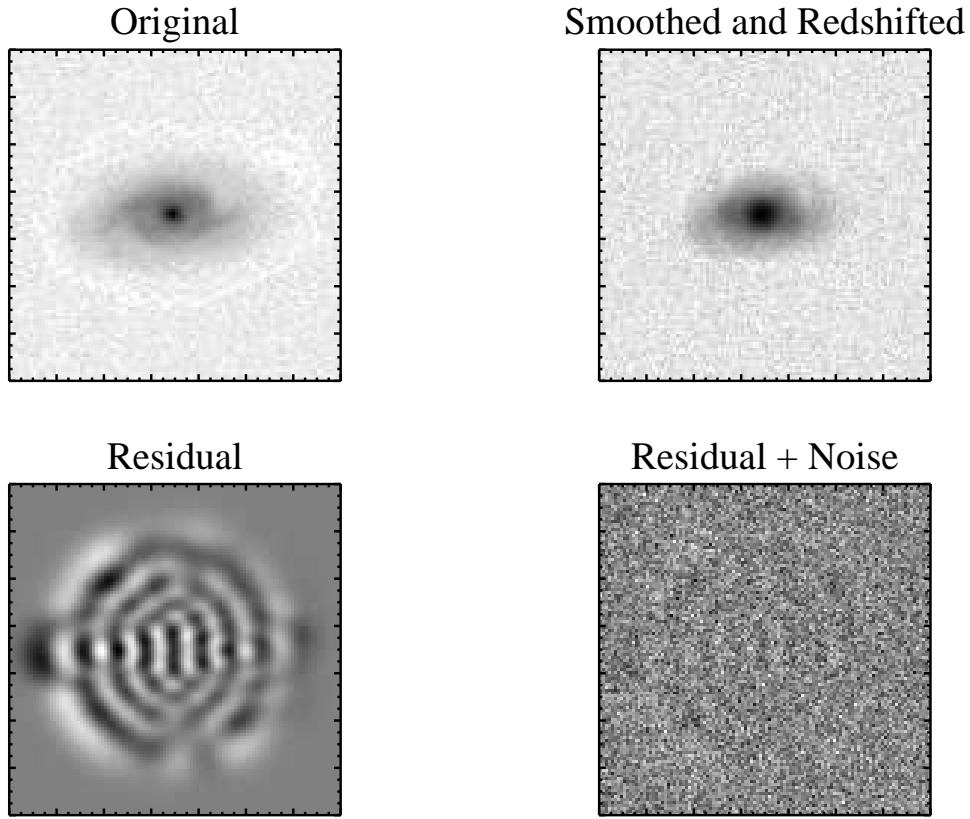}}}
\caption{Clockwise from the upper left: Original $r$-band Image of a galaxy, the galaxy image after artificial redshifting and smoothing to a Gaussian PSF of $FWHM = 2.0$ kpc, the residual of the smoothed and redshifted galaxy image and that reconstructed from its elliptical shapelet coefficients, and the residual plus noise. The residual is well below the noise level. The square root of the two galaxy images is shown to bring out the low surface brightness features, and we have added artificial noise to the smoother galaxy image in order to make it more comparable to the original image. \label{decomp_im}}
\end{center}
\end{figure}

After the smoothing the image, we calculate the shapelet coefficients by decomposing the galaxy image about its centroid in each band.  We compute the centroid of the image from the shapelet coefficients (see Paper I) and decompose the galaxy about this new centroid.  The centroid is defined as the first moment of the galaxy. We iterate this procedure twice.  We then calculate the total number of instrumental counts in the image in each band from the shapelet coefficients, convert this to a flux in Jy, and normalize the coefficients so that the flux of the galaxy summed over all bands is 1 Jy.  This keeps information regarding the differences in a galaxy's flux between bands and ensures that galaxies of different total optical luminosities do not have different shapelet coefficients; i.e., only information regarding a galaxy's morphology and its flux ratios between the bands $u, g, r, i, $ and $z$ are included in the shapelet coefficients, making the resulting classification scheme independent of abolute magnitude. Figure \ref{decomp_im} shows a galaxy image at a few stages in the decomposition process.

\section{PRINCIPAL COMPONENT ANALYSIS}

\label{pca}

\subsection{The Transform}

\label{transform}

In order to reduce the dimensionality of our data set, we perform a principal component analysis (KL-transform) on the shapelet coefficients for the galaxies of Sample 2.  Doing so allows us to reduce the 455-dimensional space spanned by the shapelet coefficients (91 coefficients for each band) to one that is more manageable.  The principal component analysis was performed on the `sum-of-squares and cross-products' (SSCP) of the data matrix \citep[e.g.,][]{murtagh}, where the data matrix is constructed by concatenating the shapelets coefficients in each band into one matrix that contains the entire multiwavelength shapelet information.  In other words, the KL-transform was done using the entire multi-band information. We removed data points outside a distance of $10\sigma$ from the median for each shapelet coefficient before calculating the SSCP matrix, as we assumed that either data points outside of this range were the result of some sort of error in the reduction processes and thus unphysical, or would result in principal components that appear to have large variance due to the presence outlying points. This resulted in removing $\sim 10\%$ of our original sample. We note that these galaxies were only removed from the principal component calculation; we still calculated their projections onto the principal component space and used them in the mixture model classification. A brief look did not reveal any obvious visual differences between these galaxies and those used in calculating the SSCP matrix. We chose to do the PCA on the SSCP matrix for ease of interpretation.  The SSCP matrix is the uncentered covariance matrix, and the SSCP results differ only by a constant from that obtained by performing the PCA on the covariance matrix. The KL-modes of the SSCP PCA are nearly indistinguishable from the covariance PCA, however we prefer the SSCP KL-modes as it allows us to view the first principal component as a type of `starting point' for galaxy morphology, with subsequent modifications from the other eigenmodes resulting in a galaxy's unique morphology. For principal components obtained from the covariance matrix one must add the mean back in to reconstruct a galaxy's morphology, and in this case the first eigenmode can have positive and negative coefficients, whereas the coefficients are always positive for the SSCP result. This is important bacause it is not very helpful to view the first principal component as a starting point if its flux is negative. In summary, we prefer the SSCP results because of it's simpler interpretation as the first principal component being the `basic' galaxy morphology, as opposed to the first principal component of the covariance matrix being the `basic' galaxy morphology after adding the mean back in. Also, we do not consider performing the PCA on the correlation matrix, as it would require us to standardize the shapelet coefficients, destroying our flux normalization.

Similar to the results from Paper I, which applied the KL-transform to just the $r$-band images, the first few principal components contain the vast majority of the variance.  We denote the $j^{th}$ principal component as $v_j$ and its corresponding coefficient as $a_j$.

\begin{figure}
\begin{center}
\scalebox{0.7}{\rotatebox{0}{\plotone{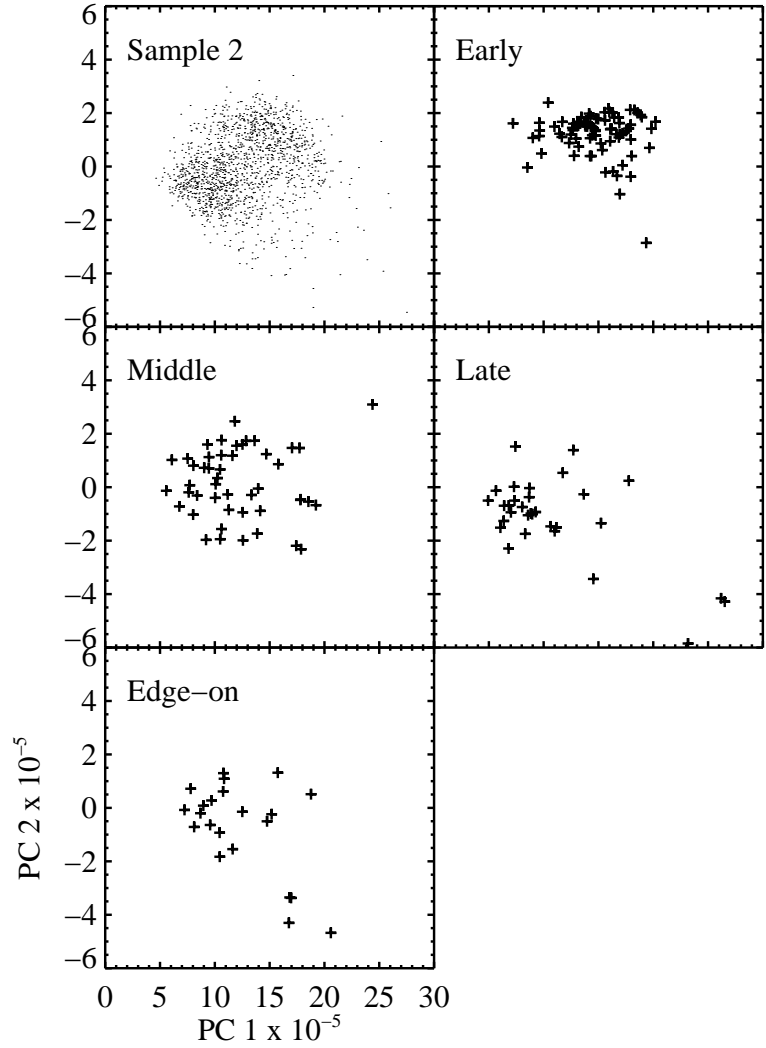}}}
\caption{Locations of the galaxies of Sample 2 (top left plot) and Sample 1 in the 2-dimensional slice spanned by $v_1$ and $v_2$.  Using the reference galaxies from Sample 1, we are able to see that the different Hubble types are well separated in this plane. \label{ttype2d}}
\end{center}
\end{figure}

\begin{figure}
\begin{center}
\scalebox{0.7}{\rotatebox{90}{\plotone{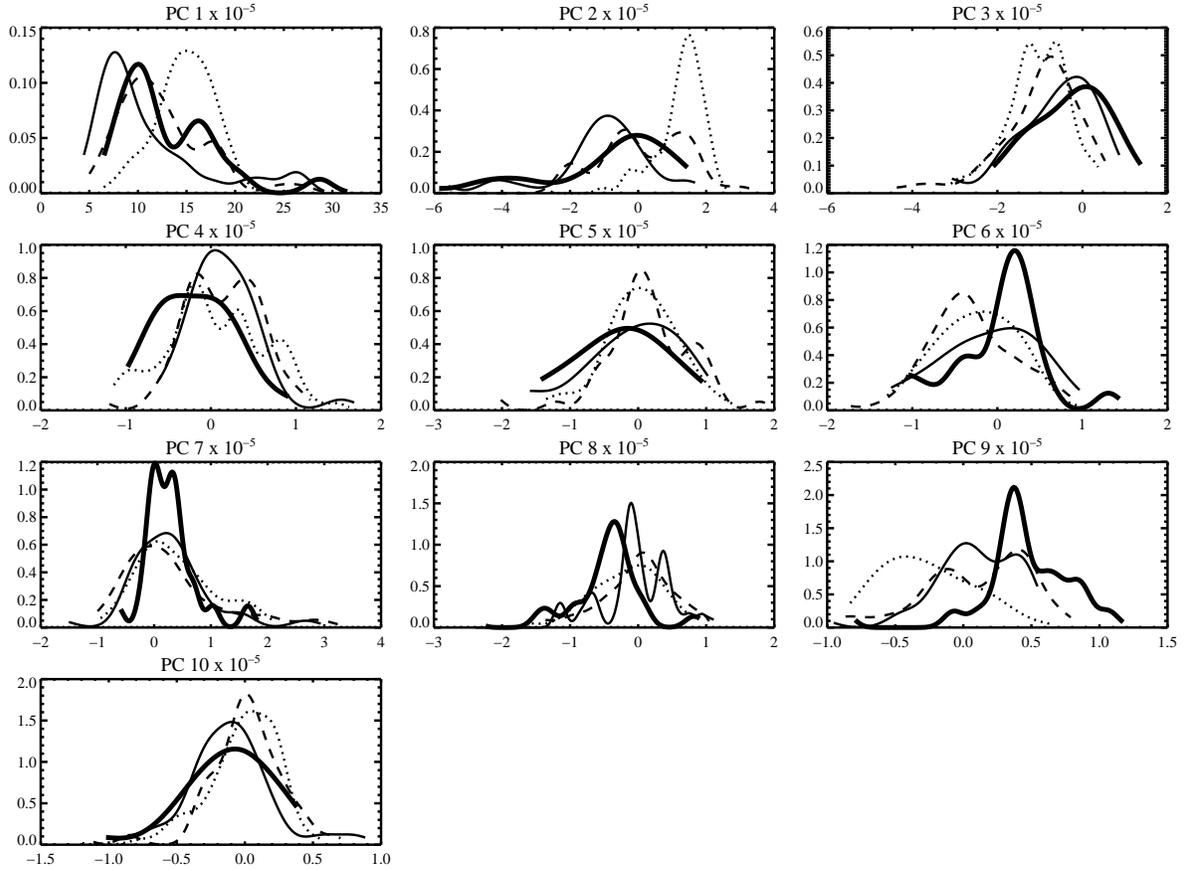}}}
\caption{Marginal probability densities of the Sample 1 galaxies in the first ten principal components. The dotted line represents early types, the dashed line middle types, the thin solid line late types, and the thick solid line edge-on spirals. From these plots, one can see that the Hubble types are well separated along a few of the $v_j$. \label{ttype1d}}
\end{center}
\end{figure}

After applying the KL-transform on the Sample 2 data, we calculate the projections of the galaxies of Sample 1 along the principal components.  Doing so allows us to develop an idea of the location of the different Hubble types in the KL-space.  We divide the Sample 1 galaxies into four types:  early, middle, late, and edge-on.  The early types include those classified by \citet{nak03} as Hubble types E--Sa, the middle types as Sab--Sbc, and the late types as Sc--Im.  The edge-on class consists of middle and late type galaxies with axis ratios $b/a < 0.4$. Figure \ref{ttype2d} shows the locations of the galaxies in the 2-dimensional slice spanned by $v_1$ and $v_2$, and Figure \ref{ttype1d} shows the marginal probability densities of the first ten $a_j$ for the Sample 1 galaxies, divided according to Hubble type.  Note that in general, the distribution of the edge-on spirals is not significantly different than that of the middle and late-type spirals, justifying our use of elliptical shapelets to remove axis ratio information to first order. The first, second, and ninth $v_j$ obtain the most significant separation of ellipticals and spirals, and to a lesser extent the third as well. Figure \ref{klbw} shows the first ten KL-morphologies, constructed from their shapelet coefficients, as well as an early and late type galaxy, shown as a reference to the relative scale of the $v_j$. For a few of the KL-morphologies, the peak does not exactly fall in the center of the image; this is a result of our decomposition method as we defined the centriod of decomposition to be the first moment of the galaxy and not the location of the flux peak. In Table \ref{tab01} we show the ratios of the total flux for each of the first ten $v_j$ in each band, as well as the ratios of the total `energy' in each band.  We define total energy in the usual mathematical way as
\begin{equation}
\int_{-\infty}^{\infty} |v_j^{(k)} ({\bf x})|^2 d^2 x = \sum_{\bf n} |f^{(k)}_{\bf n}|^2.
\label{eq09}
\end{equation}
Here, $v_j^{(k)} ({\bf x})$ is the $j^{th}$ principal component, represented in the pixel space of the original galaxy images (i.e., the image of the principal component as in Figure \ref{klbw}) for the $k^{th}$ band, and $\{ f^{(k)}_{\bf n} \}$ is the set of shapelet coefficients representing $v_j^{(k)} ({\bf x})$. The total flux ratios show how each $v_j$ contributes to the overall SED of the galaxy, whereas the energy ratios show for each $v_j$ the relative importance of each band's spectro-morphological contribution. The flux ratios are normalized such that the sum of their absolute values is unity, and the energy ratios are normalized to sum to unity. Figure \ref{klgri} shows false color images of the energy of each of the first ten $v_j$, constructed from the $g, r,$ and $i$ band images. Here, as well as in the rest of this work, we use the asinh stretch of \citet{lupt04} to display Red-Green-Blue (RGB) images. These images are helpful in visualizing the color gradient of each $v_j$, giving a spatially-dependent description of color.

\subsection{Description of the Principal Components}

\label{descript}

The first principal component is very similar to $v_1$ found from using just the $r$-band images in Paper I.  In all five bands it has a radial profile that is between an exponential and a Gaussian, and its morphology does not change significantly between the different bands.  Figure \ref{sersic} shows the azimuthally-averaged radial profile for $v_1$, as well as the best-fit S\'{e}rsic profile, defined as
\begin{equation}
I(r) = I_0 e^{-(r / r_0)^{-1/n}},
\label{eq10}
\end{equation}
where $r_0$ is the characteristic radius and $n$ the S\'{e}rsic index. The first principal component has S\'{e}rsic induces of $n \approx 0.73$, with minimal variance across the different bands.  In addition, Figure \ref{sersic} shows the $u - r$ radial profile for $v_1$; there is minimal radial color gradient in $v_1$ until one reaches the edge..

The vast majority of the variance is contained within $v_1$, and the magnitudes of its coefficients, $|a_1|$, are significantly higher than the other $|a_j|$.  This implies that $v_1$ may be interpreted as the basic galaxy morphology, where further and comparitively small modifications introduced from the other $v_j$ serve to form a galaxy's unique shape. In other words, galaxies appear to be constructed starting with a $v_1$-like component, which is then modified with additional components. Furthermore, $v_1$ does not contain any holes of negative flux, nor are any of the coefficients negative, as would be expected if $v_1$ forms the `basic' galaxy morphology.  In general, $v_1$ has astronomical color that is `redder' than the average for our sample, with the exception of $i-z$, which is $-1.42$ for $v_1$ and $0.28$ for the Sample 2 galaxies.  This KL-morphology has $u-r = 2.91$, whereas the Sample 2 average is $\langle u - r \rangle = 2.34$. In addition, as can be seen from Figure \ref{ttype1d}, the Hubble types separate very well along $v_1$, with the late types having the smallest values of $a_1$ and the early types having the largest.

The scatterplot matrix in Figure \ref{kl_vs_cur} shows the 2-d distributions as well as the 1-d marginal probability densities of Sample 2 for the first three $a_j$, the $r$-band concentration index, and $u - r$ color.  The concentration index is a common morphological measurement which we define the same way as in Paper I:
\begin{equation}
C = 5.0 \log_{10} \left ( \frac{r_{90}}{r_{50}} \right ).
\label{eq11}
\end{equation}
Here, $r_{90}$ and $r_{50}$ are the radii where the Petrosian ratio, $\eta$, is equal to 0.1 and 0.5, respectively.  More concentrated galaxies (e.g., early types) will have higher values of $C$. The 1-dimensional probability densities are estimated here, as well as elsewhere in this work, via kernel density estimation.  The kernel estimate takes the empirical probability density which places mass $1 / N$ at each data point, $x_i$, for $N$ data points, and convolves them with a kernel, $K(x)$, of bandwidth $h$:
\begin{equation}
p(x) = \frac{1}{N h} \sum_{i=1}^{N} K \left( \frac{x - x_i}{h} \right) .
\label{eq12}
\end{equation}
The kernel is also a probability density and is constrained to integrate to one.  It is well known that the choice of kernel is not important, and in this work we use the standard normal density. However, the choice of bandwidth, $h$, {\em is} important and there has been a significant amount of work toward finding the optimal bandwidth.  In this work we use the Sheather and Jones plug-in bandwidth \citep{sj91}. The kernel estimate provides an accurate and smooth estimate of the probabilty density of a variable, and is superior to standard histogram estimates. It is possible to deal with measurement errors in the kernel density estimate \citep[e.g.,][]{carr04}, and this involves deconvolving the observed probability density with the probability density of the errors. We did not do this, as this typically only makes a significant difference when the average variance from the measurement errors is some non-negligable fraction of the sample variance. Furthermore, it is unlikely that accounting for the measurement errors would significantly effect our discussion in \S~\ref{mix}, and we choose to take the simpler path of neglecting the measurement errors because a robust investigation of the probability densities of the various physical parameters is not our goal.

In order to faciliate the subsequent discussion of the principal components, we show in Figure \ref{klmod} the results when each principal component is added to or subtracted from the first one.

The second principal component has almost half of its energy in the $g$-band, implying that the most important spectro-morphological contribution from $v_2$ is in the $g$-band. This eigenmorphology is `red' in the sense that the flux is negative for the $u$ and $g$ bands and positive for the redder bands, and it adds or subtracts a red core and blue halo.  It should be noted that the RGB images seen in Figure \ref{klgri} are of the {\em energy} of the principal components, and are thus necessarily positive. For example, while $v_2$ may appear to have a blue core in this image, this is merely a reflection of the fact that the absolute value of the bluer core flux is higher than that of the redder core flux; however, inspection of the images in Figure \ref{klbw} shows that the core flux becomes {\em negative} as ones moves to the blue. Therefore, adding this spectro-morphology results in decreasing the blue core flux and increasing the red core flux, creating a red core.

The second principal component has the largest ratio of $u$-band energy to $r$-band energy, signifying that $v_2$ gives the largest contribution to the $u - r$ color gradient of a galaxy.  The values of $a_2$ are correlated with $u - r$ color, as can be seen from Figure \ref{kl_vs_cur}.  It is interesting to note that the zero point of $a_2$ occurs at a value of $u - r \sim 2.3$.  This is very close to the optimal color separator ($u - r = 2.22$) of the bimodal distribution of galaxies in color space found from SDSS data \citep{strat01, bald03}.  Late types have negative values of $a_2$, while the early types have positive values, as would be expected from the well known color-morphology relationship.  Morphologically, positive values of $a_2$ have the effect of increasing the central concentration, and making a redder more concentrated core with a bluer, more extended halo.  Negative values do the opposite, making the core bluer and more extended, and the halo redder and of lower surface brightness.  In addition, we note that $a_2$ is also correlated with concentration index, as expected from its spectro-morphological contribution.

The third KL-morphology is dominated by negative flux in the $u$ and $g$ bands, has positive flux near the core in the $r$ and $i$ bands, and is dominated by positive flux in the $z$ band. The spectro-morphological energy is approximately evenly distributed accross the $g$, $r$, and $i$ bands. The overall spectral contribution from $v_3$ is to make the astronomical colors more red for positive $a_3$, except for $i - z$ which is made bluer.  Positive values of $a_3$ make a galaxy more red, especially near its edge, and elongate the core along the major axis. Negative $a_3$ produce a bluer core and a bluer and brighter halo, making the galaxy appear more extended.

The fourth eigenmorphology display asymmetric structure in the $g$ and $r$ bands, and to a lesser extant in the $i$ band, and a significant amount of positive flux in the $z$ band. In fact, $v_4$ has a significant fraction of is spectro-morphological information in the $z$ band, and the $z, r$ and $i$ bands contain almost all of the spectro-morphological energy.  In contrast to the previous two principal components, positive $a_4$ make the astronomical colors bluer, except for $i - z$ which is made redder.  Positive $a_4$ shift the peak of the core in the direction of positive flux in the asymmetric $g, r$, and $i$ images of $v_4$, and produce a more concentrated, bluer core and brighter, bluer halo. In addition, positive $a_4$ appear to create a small red annulus around the core, and an inflection point reminescent of a weak separation of `bulge' and `disk' components. For negative $a_j$, the core peak is shifted in the opposite direction and the core becomes more red. The bulge component is more red and the disk component receiving the largest contribution in flux from the $r$ band; the halo is made slightly redder.

The fifth $v_j$ appears to pick up asymmetry along the major axis, and its spectro-morphological energy is dominated by the $r$ and $i$ bands. As with $v_4$, positive $a_5$ make a galaxy's colors bluer, with the exception being $i - z$. Positive values make a galaxy slightly bluer, with an asymmteric and narrower core. The core is redder on the side that is more concentrated. Negative $a_5$ have the opposite effect, producing a slightly redder and less concentrated core, with the core being bluer on the more concentrated side.

The sixth eigenmorphology is similar to the fifth, with the asymmetry along the minor axis, and the spectro-morphological energy being dominated by the $r$- and $i$-band components.  As before, positive $a_6$ make a galaxy's broad-band colors slightly bluer, with the exception this time being $r - i$. This KL-morphology causes a galaxy to be more red on one side of the major axis, and more blue on the other. In addition, the core peak is shifted in the direction of the bluer flux.

The seventh principal component is very similar to the fourth but with opposite-signed flux.  Also, the $i$ band is less important for $v_7$ than it was for $v_4$, with the $z$ band being more important. In fact, $v_7$ makes the greatest contribution to the $z$-band out of the first twelve $v_j$.  There are no broad trends in the contribution of $v_7$ to the astronomical colors as the previous six KL-modes, but positive $a_7$ make $u - r$ decrease while negative $a_7$ increase it.  Positive $a_7$ shift the peak, make the galaxy more asymmetric and less concentrated,  and create a red central core within a bluer envelope. Negative $a_j$ have the opposite effect, making the galaxy more concentrated with a bluer central core inside of a red envelope and a faint blue halo.

The eighth KL-morphology has a narrow central core, surrounded by a ring of negative flux. The flux is negative along the major axis and positive along the minor, and the spectral morphology is dominated by the $r$ and $i$ bands. Positive $a_8$ slightly decrease $u - r$. Positive $a_8$ create a narrower, bluer core surrounded by a red annulus. The flux becomes red and less extended along the major axis, whereas a blue halo extends along the minor axis. The red annular region corresponds to an inflection point in the major-axis profile, hinting at a bulge-disk separation. Negative $a_8$, on the other hand, create a broader, bluer core with a flatter red peak.  The peak is particularly flat along the minor axis. A blue halo extends along the major axis like that along the minor axis for positive $a_8$.

The ninth $v_j$ is dominated by the $r$- and $i$-band spectro-morphological energy, with a greater contribution from the $g$-band than most $v_j$. Positive $a_9$ decrease the $u-r$ color. This KL-morphology has a narrow core of positive flux, followed by a region of negative flux, followed by another region of positive flux.  This $v_j$ makes an important contribution to the morphology of spiral galaxies, as positive $a_9$ contribute strongly to separating a `bulge' and `disk' component as well as creating spiral arms. Positive $a_9$ create a narrow, slightly bluer core, with a dip in the radial profile in the region of red flux, and a blue region that is initially brighter than the red region and gradually dims.  The red region corresponds to the region in galaxies between the central core and the spiral arm, where the flux is from the red bulge. Spiral arms typically extend away from the core at an angle and sweep around to cross the major axis at a larger radius; this creates a gap in the flux along the major axis where the flux is dominated by red light from the bulge stars. Negative $a_9$ create a slightly bluer core with a slightly redder center.  The core is morphologically more `boxy', and is surrounded by a red halo.

The tenth principal component is dominated by its $r$-band spectral morphology. This component's contribution is spectral, as $v_{10}$ does not produce any noticeable changes to the morphology of $v_1$. Positive $a_{10}$ create a significantly bluer version of $v_1$ and negative create a significantly redder version. This KL-morphology is predominately the result of several outlying galaxies with unusual $r$-band flux that dominate the variance in this $v_j$.

It should be noted that the description of the principal components given here is largely based on each principal component's contribution {\em independent} of the others.  While this is helpful for interpreting the individual KL-morphologies, one should be careful when analyzing the $a_j$ of galaxies, as the joint probability distribution of the entire set of $a_j$ must be taken into account. For example, simply because a galaxy has a positive value for $a_2$, this does not necessarily mean that it will have a concentrated red core; indeed, it may be that if $a_2$ has this value then it is more likely that the other $a_j$ will have values such that the final spectral morphology is that of, say, a blue broad core. That the entire joint distribution of the $a_j$ must be taken into accont can be seen in the preceeding discussion regarding the higher values of $a_1$ for early types. If one were to predict a galaxy's spectral morphology from $a_1$ independent of the remaining $a_j$, then one would conclude that early types are {\em less} concentrated than late types. This is not true of course. In fact, if one takes into account the joint probability density of the $a_j$ one would see that if a galaxy has a large value of $a_1$, then it is likely to have certain values of the other $a_j$ that result in a {\em more} concentrated core than that of galaxies with low values of $a_1$. In general, it appears that the joint distributions are approximately independent for all $a_j, j > 2$, with the exception of $a_9$, and the preceding discussion should provide a useful guide in interpreting a galaxy's $a_j$.

\section{CLASSIFICATION}

\label{the_method}

\label{classification}

We use the {\bf fastem}\footnote{{\bf fastem} is an improvement upon the {\bf fastmix} software we used in Paper I.} software \citep{moo99, conn} developed by the Auton Lab at Carnegie Mellon University to estimate the probability density of the Sample 2 data in the 2-dimensional space spanned by the first two principal components.  The density is modeled as a mixture of 2-dimensional Gaussians, with each of these Gaussians representing a different class.  We only use the first two principal components because the density becomes too sparse in higher dimensions and the algorithm returns a single Gaussian in these cases. A large sample of galaxies would enable mixture model classification in a higher dimensional space. We find that the density is best fit with three Gaussians, where we use the Bayesian Information Criteria ($BIC$) to select the number of Gaussians.  The $BIC$ is defined in several ways, here we use the form
\begin{equation}
BIC = -2 \ell (\theta) + d \log N,
\label{eq13}
\end{equation}
where $\ell (\theta)$ is the log-likelihood of the model with parameters $\theta$, $N$ is the number of data points, and $d$ is the number of parameters. Minimizing $BIC$ is approximately equivalent to choosing the model with the largest posterior probability.  Furthermore, using the $BIC$ allows us compare the relative posterior probability, $P({\cal M}_m | Z)$, of the $m^{th}$ model, ${\cal M}_m$, conditional on the training data, $Z$ :
\begin{equation}
P({\cal M}_m | Z) = \frac{e^{-(1/2) BIC_m}}{\sum_{l=1}^{M} e^{-(1/2) BIC_l}}.
\label{eq14}
\end{equation}
Here $M$ is the number of candidate models. Using Equation (\ref{eq14}), we find that the three Gaussian mixture model is much more likely than the two and four Gaussian model (as much as $\sim e^{25}$--$e^{80}$ times more likely), conditional on the data.

Paper I describes in further detail our procedure and motivation for using a mixture of Gaussians model; here we present only a description of the results.  We will use the notation $M_k$ to denote the $k^{th}$ mixture class.  Figure \ref{dec_bounds} shows the 2-dimensional joint probability density, $p(a_1, a_2)$, estimated from the mixture model fit, along with the decision boundaries used in the classification. For comparison we also show the decision boundary separating `red' and `blue' galaxies in this space, where we use the commonly used decision boundary at $u - r = 2.22$ in the red/blue classifcation. The decision boundary in the 2-dimensional KL-space spanned by $\{v_1, v_2 \}$ was estimated using Quadratic Discriminant Analysis (QDA). QDA is a commonly used method of classification that assumes each class has a Gaussian probability density.  This is the same motivation as in the mixture model, however in this case the data have been classified (i.e., red or blue), so we fit the probability density of each class separately by estimating the Gaussian parameters for that class, i.e, the covariance matrix, mean, and weight.  In reality, the red and blue classes are not normally distributed, however they are approximately so and modeling them as such will only introduce a small bias in the decision boundary estimation. We compare the two classification methods in \S~\ref{mix_vs_ur}.

Figure \ref{mixdens1d} shows the marginal probability densities for the $a_j$ of the $k^{th}$ mixture class, $p_k(a_j)$. The marginal probability densities for the first two $v_j$ were taken directly from the mixture model fit, the others were estimated by kernel density estimation in the same manner as described earlier. For the purpose of plotting and estimating $p_k (a_j)$, galaxies are taken to occupy the class of highest probability; however, it should be noted that because this classification scheme is continuous, each galaxy has a probability of occupying each class. Typically galaxies were robustly classified, with only $4\%$ having a probability of $\leq 0.6$ and $80\%$ having a probability of $\geq 0.9$ of occupying their assigned class. Here, as well as elsewhere in this work, we calculate the mean values for each class using all of the Sample 2 galaxies, weighting each galaxy by its probabiliy of being in that class. In addition, Figure \ref{ttype_vs_mix} compares the mixture classes with Hubble type, where the $a_j$ for the galaxies of Sample 1 were used to assign a mixture class based on the fit to the Sample 2 data. Figure \ref{avgmix} shows the mean morphology of each mixture class, where the means are calculated over all galaxies in Sample 2, weighted by the probability the galaxy is in each respective class. Figure \ref{mixgals} shows an individual galaxy from each $M_k$ that lies close to the mean vector for that class, and Figure \ref{phymix} shows the marginal probability densities for several physical parameters for each class, estimated with the kernel method. Only 423 of the galaxies had velocity dispersion data, and of these only a few were $M_1$ galaxies. Because of this we did not estimate the velocity dispersion density for $M_1$.

\subsection{Description of the Classes}

\label{mix}

The third mixture class, $M_3$, is dominated by early type galaxies, i.e., ellipticals and some early spirals. The average Hubble type for $M_3$ is in between S0 and Sa, and $83\%$ of $M_3$ galaxies are E/S0/Sa as estimated from the Sample 1 galaxies. Based on the mixture model fit to the Sample 2 data, we estimate that $49.3\%$ of galaxies within the limits of our sample are of class $M_3$. Galaxies in this mixture class tend to be concentrated and red, and this is certainly the case for the mean morphology of $M_3$. In general, the mean and mode of $a_j$ does not differ significantly from that of the second (spiral) mixture class, with the execptions being $a_1$, $a_2$, and $a_9$. These three KL-morphologies play the most significant role out of the first ten in separating the spectro-morphological properties of spirals and ellipticals, and it makes sense that the means and modes of these $a_j$ would differ the most noticeably between an early type class and a late type class. With the exception of the two asymmetry eigenmorphologies, $v_5$ and $v_6$, the marginal probability densities of the remaining $a_j$ ($a_3, a_4, a_7, a_8$ and $a_{10}$) tend to be broader and more skewed for $M_3$ as compared to the late type class, $M_2$. The $v_j$ for which the $p_3 (a_j)$ are significantly asymmetric tend to contribute to the central concentration, however it is not exactly clear how to interpret their joint probability density within the context of a galaxy's spectral morphology. Galaxies in $M_3$ appear to have no preference for major or minor axis asymmetry, as their $a_5$ and $a_6$ are symmetrically distributed about zero. 

The physical parameters and non-KL measures of morphology and SED are consistent with the assignment of $M_3$ to early types.  Galaxies in $M_3$ tend to have higher velocity dispersions, are slightly brighter in both $u$ and $r$, have redder $u - r$ color and spectral eigenclass, are physically smaller, more concentrated, and have higher surface brightness as compared to the late type class. The SDSS spectral eigenclass \citep{yip} is a spectral classification based on a principal component analysis of 170,000 Sloan galaxies.  Negative values correspond to `red' ($u-r \gtrsim 2.2$) galaxies, positive values correspond to `blue' galaxies ($u-r \lesssim 2.2$). The probability densities of $u-r$ and spectral eigenclass for $M_3$ are narrower than that of the spiral class, $M_2$, but with long tails that extend off in the blue directions.  This implies that there is a population of galaxies with elliptical morphologies but blue colors; we discuss this further in \S~\ref{mix_vs_ur} and \S~\ref{conclusions}. It may be that this subclass of `blue ellipticals' are the primary cause for the skewness observed in the marginal probability densities of $a_3, a_4, a_7, a_8,$ and $a_{10}$.

The second mixture class, $M_2$, has a mean Hubble type of in between Sb and Sc, is dominated by late type galaxies and contains $46.7\%$ of galaxies. Galaxies with Hubble types between Sb and Sdm are predominantly of $M_2$, and based on the Sample 1 data, we estimate that $90\%$ of $M_2$ galaxies are of Hubble types Sb and later. Galaxies in $M_2$ tend to have yellow bulges with blue spiral arms, as inferred from the mean morphology of $M_2$. The blue spiral arms in the mean spectral morphology have been `averaged out' over the many galaxies of $M_2$, and even among the individual galaxies of $M_2$ the resolution in the shapelet decomposition is such that the arms typically appear blurry and are not well-defined. However, a faint blue ring is present along the outside of the yellow bulge. This mixture class has the lowest values of $a_1$, allowing the relative contributions from the other principal components to be stronger, resulting in a more complex morphology. Furthermore, the mean and mode for $a_2$ and $a_9$ are significantly different from that of the early type class, making galaxies in $M_2$ bluer and having morphological features consistent with spiral galaxies. The other $a_j$ are approximately symmetrically distributed about zero, and in general their probability density is not heavily skewed or irregularly shaped and appears to be the narrowest.

The physical parameters and non-KL measures of morphology and SED are consistent with $M_2$ being populated by spirals. Galaxies in $M_2$ have lower velocity dispersions, bluer $u-r$ color and spectral eigenclass, are slightly dimmer than $M_3$, physically larger, less concentrated, and have lower surface brightness.  Also, the nearly uniform probability density of axis ratio means that we are just as likely to find a face-on spiral as an edge-on in $M_2$, implying that we have succeeded in removing the ellipticity information, at least to first order, through the use of elliptical shapelets. This mixture class has a broader distribution in $u-r$ and spectral eigenclass than the other two, and has a tail that extends off in the red direction of these two spectral parameters. However, this red tail is not as distinct as the blue tail of $M_3$ and the red tail of $M_1$.

The first mixture class, $M_1$, contains only about $4\%$ of galaxies within the redshift and luminosity range of Sample 2 and may perhaps be the most interesting.  Galaxies of $M_1$ appear to be nearly uniformly distributed in Hubble type, however they are the most common for Hubble types Sdm and Im. The mean morphology for $M_1$ is that of a blue strong bulge surrounded by a faint red halo. The $u$-band flux is noticeably stronger in the mean morphology for $M_1$ than for the other two classes. The joint probability density of the $a_j$ for $M_1$ is the broadest and most irregular of the three classes, and the mean and mode of most of the $a_j$ are noticeably nonzero.  In particular, the distinctly nonzero means of $a_2, a_3, a_8, a_9,$ and $a_{10}$ for this class are consistent with galaxies in this class showing a preference for a blue, less concentrated bulge surrounded by a faint red halo. In addition, the nonzero mode of $a_5$ and the bimodal distibution of $a_6$ imply that galaxies in $M_1$ tend to be asymmetric, but it is not clear to us exactly why the marginal density of $a_5$ is not bimodal as well. The noticeable preference for positive $a_{10}$ implies that galaxies that are classified as belonging to $M_1$ have abnormally low $r$-band flux.

Galaxies in $M_1$ tend to have very blue values of $u-r$ color and spectral eigenclass, are dimmer in the $r$-band than the other two classes, are of small to medium physical size, slightly more concentrated than the galaxies of $M_2$, and have high surface brightness in the $u$-band. The distribution of spectral eigenclass for $M_1$ has an extremely long tail the extends into the red direction.

\subsection{Comparison with $u-r$ Classification}

\label{mix_vs_ur}

It is interesting to compare our classification with the simple separation of `red' and `blue' galaxies typically done on SDSS data with a cut in $u-r$.  We show in Figure \ref{ur_vs_mix} the total probability density of $u - r$, as well as the density of $u-r$ for each mixture class, scaled according to their relative prevalences.  In Figure \ref{odd_urgals} we show a galaxy from each of the $M_k$ with a value of $u - r$ color unexpected for that class; e.g., because $M_3$ would be considered a red class (on average, $u - r > 2.22$), we show a galaxy from $M_3$ with $u - r$ on the blue side of the optimal color separator ($u - r = 2.22$).  A significantly larger fraction of the galaxies for $M_2$ and $M_3$ with uncharacteristic $u-r$ values for their respective class had probabilities of being in that class of $\lesssim 0.9$, implying that more of these uncharacteristic galaxies have spectro-morphological properties distinctive of more than one class.  Each of the galaxies we show in Figure \ref{odd_urgals} has a probability of being in their class of $\ge 0.972$, so we can be confident in their classification.

From the images shown in Figure \ref{odd_urgals}, we can see that our classification scheme is able to group galaxies of similar spectral morphology into different classes even when such galaxies exhibit broad-band color more characteristic of another class.  For example, the edge-on spiral of $M_2$, is very red and has a value of $u - r$ that would be expected for ellipticals, however the mixture classification is able to effectively place this galaxy into its appropriate spectro-morphological class by incorporating the morphological information. Furthermore, the unusually blue galaxy of $M_3$ has an obvious early type morphology, but one can see from the RGB image that this galaxy is bluer than is typical for this class. Similarly, the red galaxy of $M_1$ has the usual extended, less concentrated bulge, but with uncharacteristicly high $r$-band flux.

Figures \ref{dec_bounds} and \ref{ur_vs_mix} support the discussion given above. One can see from these figures that if the goal is to separate early and late type galaxies, then the $u-r = 2.22$ decision boundary is effective but not optimal. In fact, a considerable number of late type galaxies are misclassified using the $u-r$ decision boundary, and the probability densities for the $M_k$ shown in Figure \ref{ur_vs_mix} seem to suggest an optimal decision boundary at $u-r \sim 2.4$. Traditionally, late type morphologies are associated with the blue class and early with the red, and the blue and red distributions are both assumed to be Gaussian \citep{bald03}. However, the results here suggest that the $u-r$ probability densities of late and early type morphologies are only approximately normal and exhibit tails that extend past the $u-r=2.22$ decision boundary.  If we equate $M_2$ with late type galaxies, class $M_3$ with early type galaxies, and ignore $M_1$ galaxies, then the $u-r=2.22$ decision boundary only classifies $68.7\%$ of late type galaxies correctly and $88.4\%$ of early type galaxies correctly, with an average correct classification rate of $78.8\%$. This is similar to the results of \citet{strat01}, who used a sample of 287 galaxies morphologically classified by eye. They found that the $u-r=2.22$ decision boundary correctly classified $66\%$ of late types and $80\%$ of early types. In constract, using a decision boundary at $u-r=2.4$, as suggested by the probability densities in Figure \ref{ur_vs_mix}, results in $82.6\%$ of late type galaxies being classified correctly and $81.7\%$ of early type galaxies being classified correctly.  The average correct classification rate for the $u-r=2.4$ decision boundary is $82.1\%$ for this sample, about $4\%$ better than that of the $u-r=2.22$ decision boundary. Although the improvement in average misclassification rate is modest, the misclassification rate is balanced between the early and late type galaxies for the $u-r=2.4$ decision boundary.

We do not think that the bias and variance of the kernel estimate will significantly alter this result. The bias is defined as the difference between the expectation value of an estimate and the true value. The bias in the kernel density estimate is proportional to the square of the bandwidth multplied by the second derivative of the true density. Because of this, the bias will be large in regions of high curvature.  If we assume that the $u-r$ density estimates for $M_2$ and $M_3$ are approximately equal to the true densities, then the bias will be too small to significantly alter the location of the $u-r \sim 2.4$ decision boundary suggested by the kernel estimates, as the probability densities are almost linear near $u-r \sim 2.4$ and thus will have nearly zero curvature. Therefore, most of the contribution to the uncertainty will come from the variance and a pointwise 95\% confidence interval may be estimated from this variance. Although we do not show it in Figure \ref{ur_vs_mix}, a decision boundary between $2.3 \lesssim u-r \lesssim 2.5$ is consistent with the 95\% pointwise confidence interval of the density estimate. The standard $u-r=2.22$ decision boundary is outside of this interval.

These results are intriguing, and a more in-depth analysis of the $u-r$ probability densities would make a correction to these estimates that accounts for the measurement error in $u-r$, the dependence on luminosity, and sample selection. However, further analysis of using $u-r$ as a proxy for morphological type is beyond the scope of this paper.

Class $M_1$ is completely hidden in $u-r$, and extracting the rare $M_1$ galaxies within the redshift and luminosity range of our sample is impossible with a $u-r$ decision boundary. This is because the $u-r$ probability density of $M_1$ is always lower than that of the other two classes. However, because our sample is selected for $M_r < -19$, $M_1$ galaxies may dominate at fainter luminosities and it would be possible to classify them with a $u-r$ decision boundary.

\section{CONCLUSIONS}

\label{conclusions}

An important project of modern extragalactic astronomy is the pursuit of a quantitative and automatic morphological galaxy classification scheme that incorporates spectroscopic information.  Traditional classification is becoming more inadequate, and new systems are needed.  A quantitative and multiwavelength description of morphology will allow astronomers to give quantitative relations between a galaxy's spectral morphology and its physical parameters.  In addition, automating the classification scheme allows the full use of large astronomical databases for analyzing galaxy morphology, which will be of great benefit in investigating the physical significance of a galaxy's shape.

In this paper, we have tested the shapelet decomposition method as a quantitative and automatic description of galaxy morphology across the entire optical spectrum.  We apply the method to a sample of 1519 galaxies from the Sloan Digital Sky Survey, using the images in all five observing bands ($u, g, r, i, $and $z$), and show that galaxies of known broad Hubble type separate cleanly in shapelet space.  In addition, using the vast amount of SDSS data allows the admission of powerful statistical methods of analysis, such as principal component analysis and the mixture of Gaussians model.  Applying the principal component analysis, we give a description of each principal component's contribution to spectral morphology {\em independent} of the other principal components.  Using shapelets of ellipticity equal to that of the decomposed galaxy resulted in minimal contamination of axis ratio information in the principal components.  We show that each principal component contains unique morphological information that often varies with observing band, and that the KL-space sufficiently separates galaxies that are known to have different morphologies.

Furthermore, we apply a mixture of Gaussians model to describe the density of galaxies in the space spanned by the principal components, with each Gaussian representing a spectro-morphological class.  The mixture model fit the density to three Gaussians, implying three classes.  The two dominant classes were shown to be associated with early and late type galaxies, whereas the other class was populated by galaxies that could often not be associated with any particular Hubble type.  The rare first class was shown to consist of galaxies with extended, blue bulges, and were more typically asymmetric. In addition, we show that galaxies of different morphologies differ, on average, in their physical properties.  We compared our method with a simple cut on color and show that our method is superior for separating galaxies of different spectro-morphological properties, and suggest using a $u-r \sim 2.4$ decision boundary instead of the usual $u-r = 2.22$.

Our method is in general model-independent, objective, and automatic; the fact that the method is able to separate galaxies of different morphology and color so well is promising and attests to its efficacy. However, there are a few notable places where the methods that we have employed may be expanded and improved upon, and we conclude with a discussion of these.  First, we have assumed that the spectro-morphological probability densities for the $M_k$ are each a single Gaussian. This was motivated in Paper I by the central limit theorem: galaxies of the same spectro-morphological class have experienced similar physical events that produce a mean spectral morphology, but are also subject to smaller-scale physical processes unique to each galaxy that have the effect of producing numerous independent random perturbations to spectral morphology. Therefore, by the central limit theorem, we would expect the probability densities of each $M_k$ to be approximately Gaussian. However, it is likely that this is not the case for every galaxy in each $M_k$, which will cause the spectro-morphological probability densities to diverge from normality and have more pronounced tails.  For example, it may be that $M_1$ is not a distinct class at all, but that galaxies in $M_1$ are really the result of an extended tail in the $M_2$ probability density. Within this interpretation, the mixture model fit requires an extra Gaussian to pick up the $M_2$ tail. An extended tail may also be the reason for the `blue ellipticals' of $M_3$ discussed in \S~\ref{mix}. On the other hand, these `blue ellipticals' may also be a distinct class, but we do not have enough data to produce a 4-Gaussian fit to the density with better $BIC$ score than the current 3-Gaussian fit. Although it is possible that the classes exhibit significant tails in their probability densities, we believe that in general the central limit theorem justifies the use of a mixture of Gaussians model here, and that modeling the class densities as such introduces only a small bias in the fit. One could also introduce a uniform background density, as described in \citet{conn}.

The second and more significant limitation of our method lies in using the principal components as a basis for our classification scheme. We chose to utilize the shapelet basis as a spectro-morphological basis because it provides a simple but effective means of extracting morphological information with minimal contamination from varying galaxy size and axis ratio. In addition, we performed classification in the principal component basis primarily because of its dimension-reducing properties, as the KL eigenmodes provide the best lower dimensional linear approximation to a dataset and have maximal variance subject to being orthogonal. However, this does not mean that the KL-space will be the optimal space for classification.  To be more precise, we have assumed that the galaxy spectro-morphological probability density is a mixture of normal densities, at least in the original pixel space of the galaxy image. If this is the case, then all linear transformations of the galaxy image will preserve the normality of the respective class probability densities, i.e., a linear transform of a normally-distributed variable is still a normally-distributed variable. The shapelet transform and KL transform are both linear transforms, and that we expect the {\em total} probability density to be (approximately) a multimodal normal distribution follows. By employing the KL transform as a basis for classification, we have searched for the best lower-dimensional linear approximation to galaxy spectral morphology and classified in this space.  However, this is no reason to believe that the KL-space will achieve an optimal separation between the respective classes. A better strategy is to look for linear transforms that find spectral morphologies with multimodal probability distributions, and this can be accomplished with projection pursuit \citep{fried87, fried74, jones87}. In particular, one could search for a linear transform that maximizes the amount of statistical independence among the basis vectors.  This technique is called independent component analysis \citep[ICA,][]{hyvar}. Whereas PCA seeks to find the orthogonal basis that maximizes the variance along its components, ICA seeks to find a basis that maximizes statistical independence among it components.  This is equivalent to looking for non-Gaussian projections of the data along the basis vectors.  Because multimodal probability densities are strongly non-Gaussian, the ICA form of projection pursuit would be particularly interesting as a basis for spectro-morphological classification. In addition, because the ICA basis vectors are statistically independent, or at least are as statistically independent as possible, the joint probability distribution can be separated into the individual marginal probability distributions of the basis vectors.  This fundamental property facilitates a method of interpreting the individual spectral morphologies, similar to that performed in \S~\ref{descript}.  The principal components, on the other hand, are only uncorrelated, and thus are only statistically independent in the case of unimodal Gaussian data.

Using the principal components as a basis for classification has its advantages, namely in the area of giving an economical representation of a galaxy's spectral morphology. The KL-modes have been useful for getting a `feel' for the problems investigated here, but there are certainly more effective bases for doing spectro-morphological classification. We believe that the classification method described here shows considerable promise, and further improvement can be made by choosing a more appropriate basis for spectral-morphological classification.

\section{ACKNOWLEDGMENTS}

This work was supported by National Science Foundation grants AST-F007182, AST-0206277, and NSF grant 044327. We would like to thank Alexandre Refregier for the use of his shapelet code, and Erin Sheldon and Ben Koester for helping with computer and data access issues. We would also like to thank the anonymous referee for helpful comments that contributed to the improvement of this manuscript.

Funding for the creation and distribution of the SDSS Archive has been provided by the Alfred P. Sloan Foundation, the Participating Institutions, the National Aeronautics and Space Administration, the National Science Foundation, the U.S. Department of Energy, the Japanese Monbukagakusho, and the Max Planck Society. The SDSS Web site is http://www.sdss.org/.

The SDSS is managed by the Astrophysical Research Consortium (ARC) for the Participating Institutions. The Participating Institutions are The University of Chicago, Fermilab, the Institute for Advanced Study, the Japan Participation Group, The Johns Hopkins University, Los Alamos National Laboratory, the Max-Planck-Institute for Astronomy (MPIA), the Max-Planck-Institute for Astrophysics (MPA), New Mexico State University, University of Pittsburgh, Princeton University, the United States Naval Observatory, and the University of Washington.

\clearpage

\begin{deluxetable}{ccccccccccc}
\singlespace
\tablewidth{0pt}
\tabletypesize{\scriptsize}
\tablecaption{Flux Ratios and Energy of the KL-Morphologies. \label{tab01}}
\tablehead{
\colhead{} & \multicolumn{5}{c}{Flux Ratio} & \multicolumn{5}{c}{Total Energy} \\
\colhead{KL-mode}  & \colhead{$u$} & \colhead{$g$} & \colhead{$r$} &
\colhead{$i$} & \colhead{$z$} & \colhead{$u$} & \colhead{$g$} & \colhead{$r$} &
\colhead{$i$} & \colhead{$z$}
}
\startdata 
PC 1  &  0.022 &  0.200 &  0.320 &  0.360 &  0.097 & 0.002 & 0.155 & 0.367 & 0.443 & 0.033 \\
PC 2  & -0.056 & -0.080 &  0.264 &  0.453 &  0.148 & 0.031 & 0.460 & 0.157 & 0.313 & 0.040 \\
PC 3  & -0.060 & -0.365 & -0.312 & -0.221 &  0.042 & 0.019 & 0.299 & 0.231 & 0.349 & 0.102 \\ 
PC 4  &  0.039 &  0.179 &  0.221 & -0.185 &  0.375 & 0.005 & 0.083 & 0.203 & 0.390 & 0.320 \\ 
PC 5  &  0.044 &  0.220 &  0.284 &  0.179 &  0.273 & 0.002 & 0.167 & 0.358 & 0.430 & 0.044 \\ 
PC 6  & -0.004 & -0.076 & -0.507 & -0.297 & -0.116 & 0.001 & 0.160 & 0.332 & 0.487 & 0.020 \\ 
PC 7  &  0.039 & -0.070 &  0.112 & -0.265 &  0.514 & 0.003 & 0.110 & 0.235 & 0.291 & 0.360 \\
PC 8  &  0.061 & -0.253 &  0.198 &  0.415 &  0.072 & 0.002 & 0.162 & 0.368 & 0.437 & 0.032 \\
PC 9  &  0.044 & -0.246 & -0.364 & -0.165 &  0.181 & 0.006 & 0.190 & 0.365 & 0.382 & 0.057 \\
PC 10 &  0.095 &  0.218 & -0.281 &  0.225 &  0.181 & 0.050 & 0.114 & 0.541 & 0.170 & 0.125 \\
\enddata
\tablecomments{Flux ratios are normalized such that the sum of their absolute values over all five bands is unity for each $v_j$. The total energy is normalizied such that the sum of the energies over all five bands is unity for each $v_j$. The numbers have been rounded and so may not add up to exactly one.}
\end{deluxetable}
 
\clearpage

\clearpage

\clearpage

\clearpage

\clearpage

\clearpage
\begin{figure}
\begin{center}
\scalebox{0.7}{\rotatebox{0}{\plotone{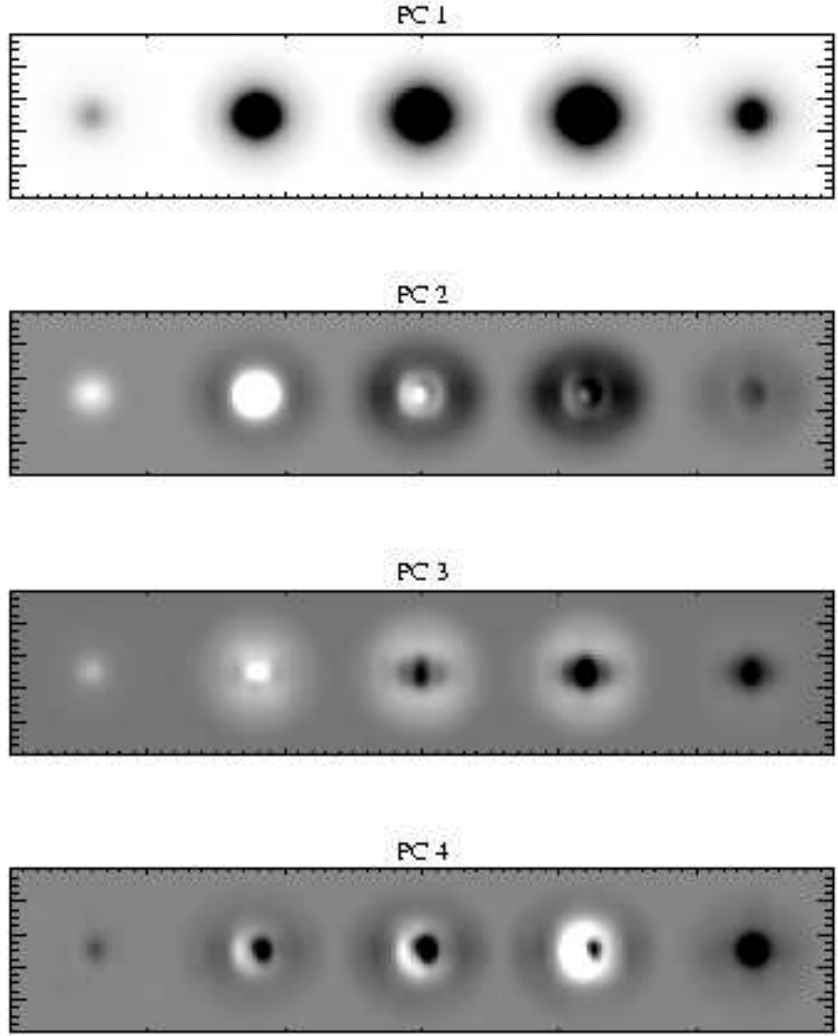}}}
\caption{Images of the first ten spectro-morphological principal components, constructed from their shapelet coefficients with an axis ratio of unity. Starting from the left, the bands of the images are: $u, g, r, i,$ and $z$. For reference, we also show an early and late type galaxy of same relative scale and similar axis ratio to the $v_j$. The images of the $v_j$ are sigma-clipped, while the galaxy images are shown with a square root stretch. For principal components two through ten, blacker areas denote positive values and whiter areas denote negative values. All pixel values of $v_1$ and the galaxy images are positive. \label{klbw}}
\end{center}
\end{figure}

\clearpage

\addtocounter{figure}{-1}

\begin{figure}
\begin{center}
\scalebox{0.8}{\rotatebox{0}{\plotone{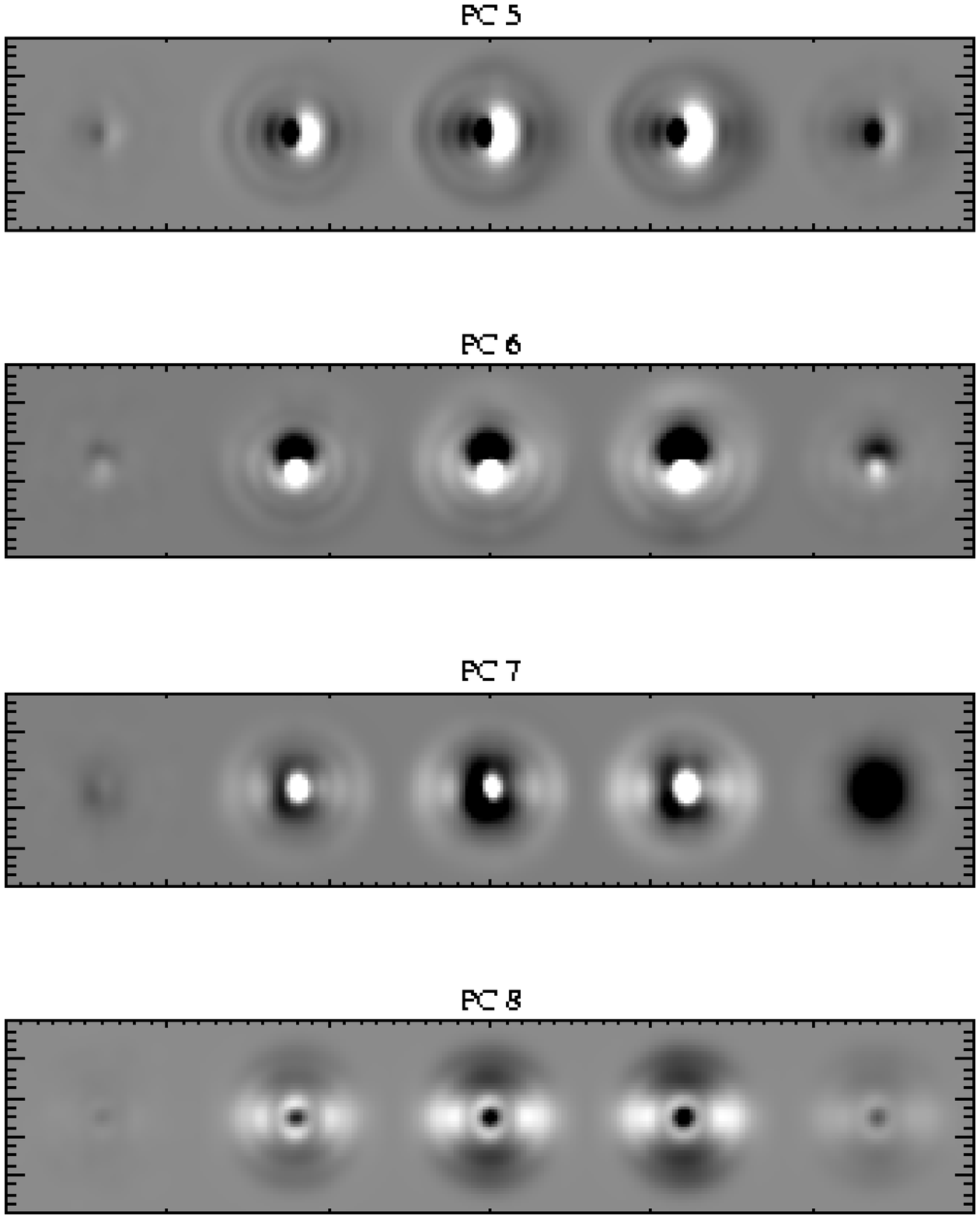}}}
\caption{cont.}
\end{center}
\end{figure}

\clearpage

\addtocounter{figure}{-1}

\begin{figure}
\begin{center}
\scalebox{0.8}{\rotatebox{0}{\plotone{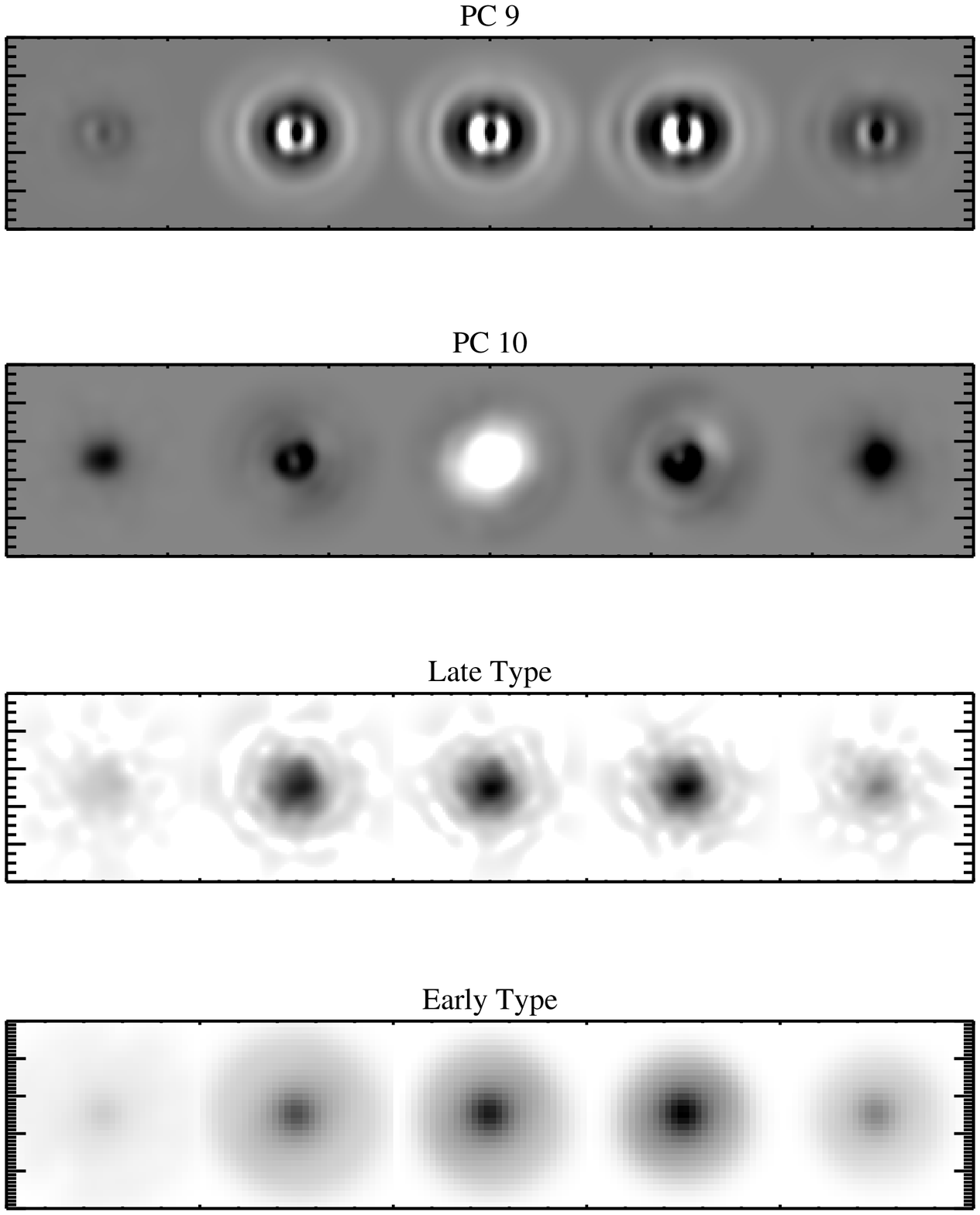}}}
\caption{cont.}
\end{center}
\end{figure}

\clearpage
\begin{figure}
\begin{center}
\scalebox{1.2}{\rotatebox{90}{\plottwo{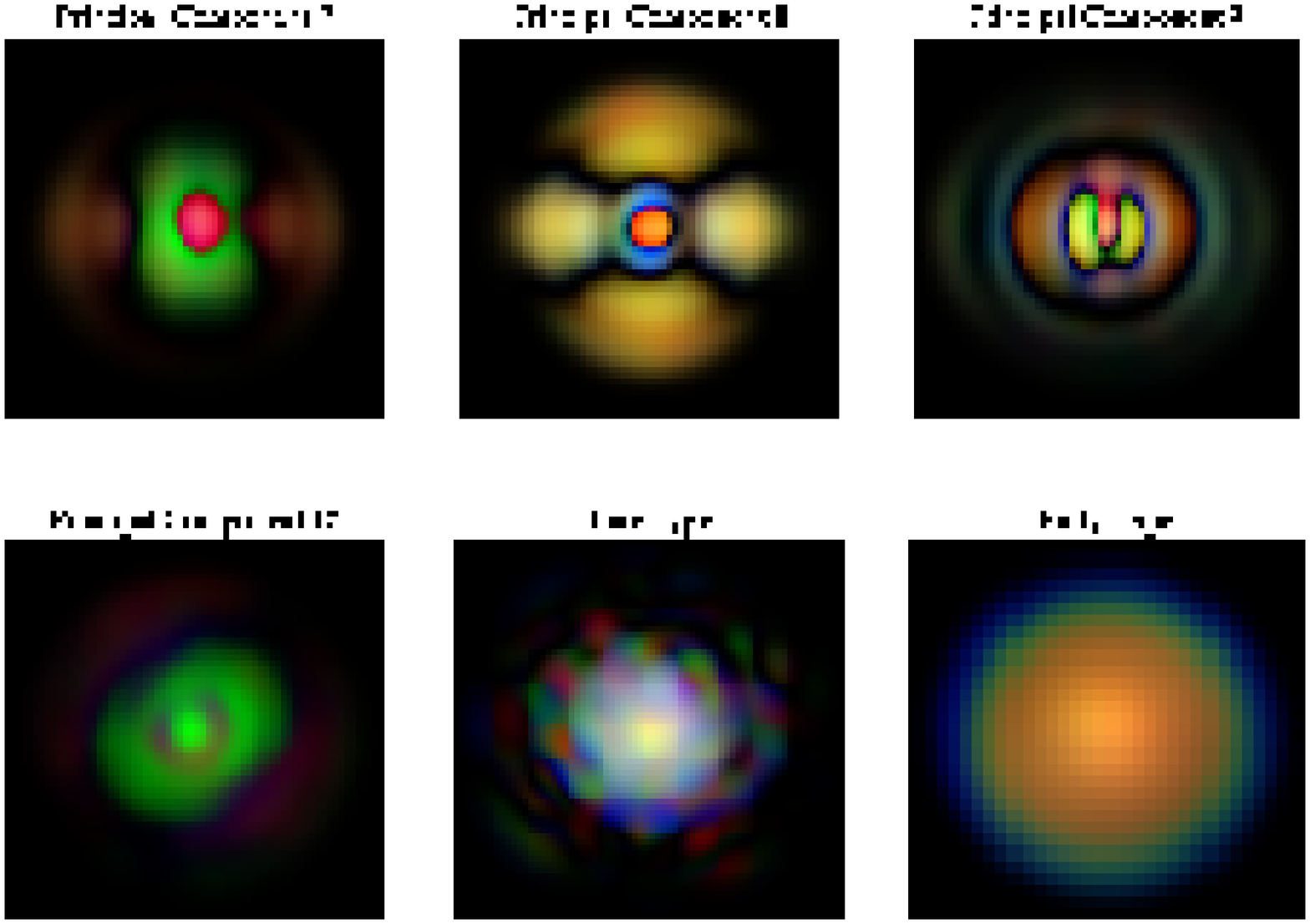}{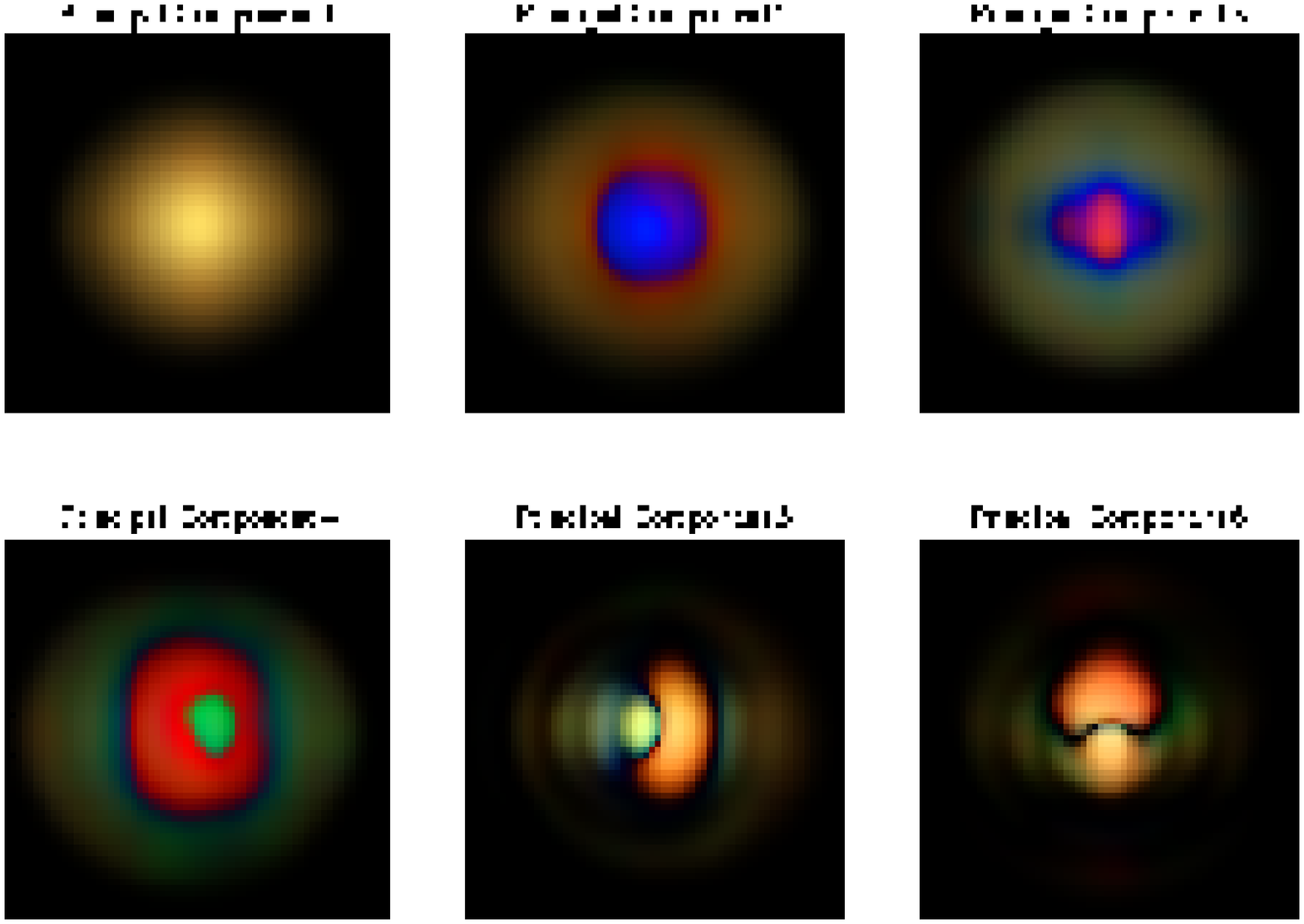}}}
\caption{False color RGB images, constructed from the $g, r,$ and $i$ band images of the energy of the first ten principal components. As described in the text, the energy is defined as the square of the image. Also shown is the $g, r$, and $i$ band energies for the same galaxies shown in Fig. \ref{klbw}. The galaxies are shown with a comparable stretch as the $v_j^2$. \label{klgri}}
\end{center}
\end{figure}

\clearpage
\begin{figure}
\begin{center}
\scalebox{0.7}{\rotatebox{90}{\plotone{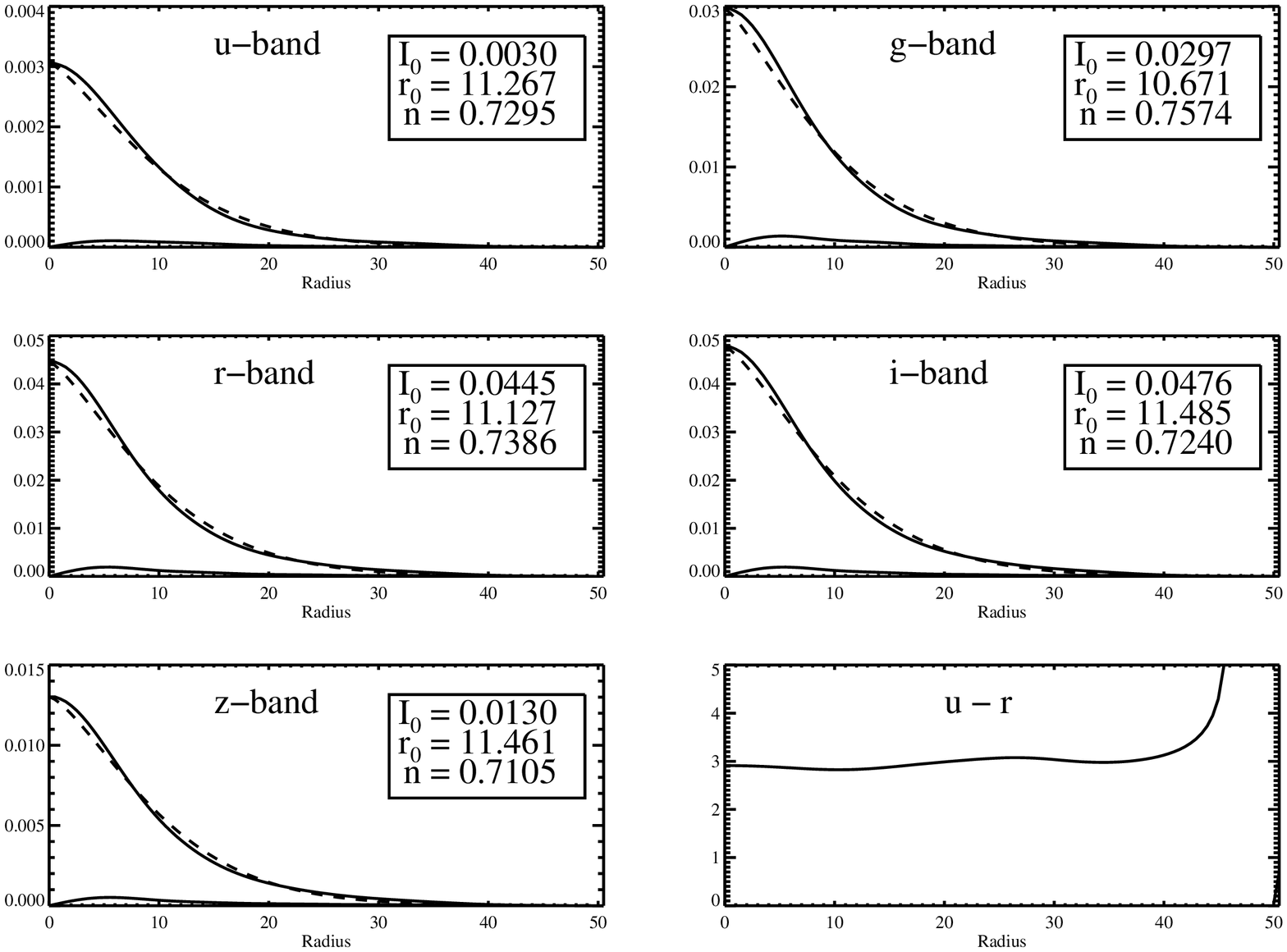}}}
\caption{The azimuthally-averaged radial profile of $v_1$ (solid line) for each band, as well as the best fit S\'{e}rsic profile (dashed line). Also shown is the azimuthally-averaged $u-r$ radial profile for $v_1$. The radius is shown in units of image pixels. Because the principal components are independent of scale, converting from pixels to physical units involves an arbitrary choice, so we just show the radius in units of pixels. \label{sersic}}
\end{center}
\end{figure}

\clearpage
\begin{figure}
\begin{center}
\scalebox{0.7}{\rotatebox{90}{\plotone{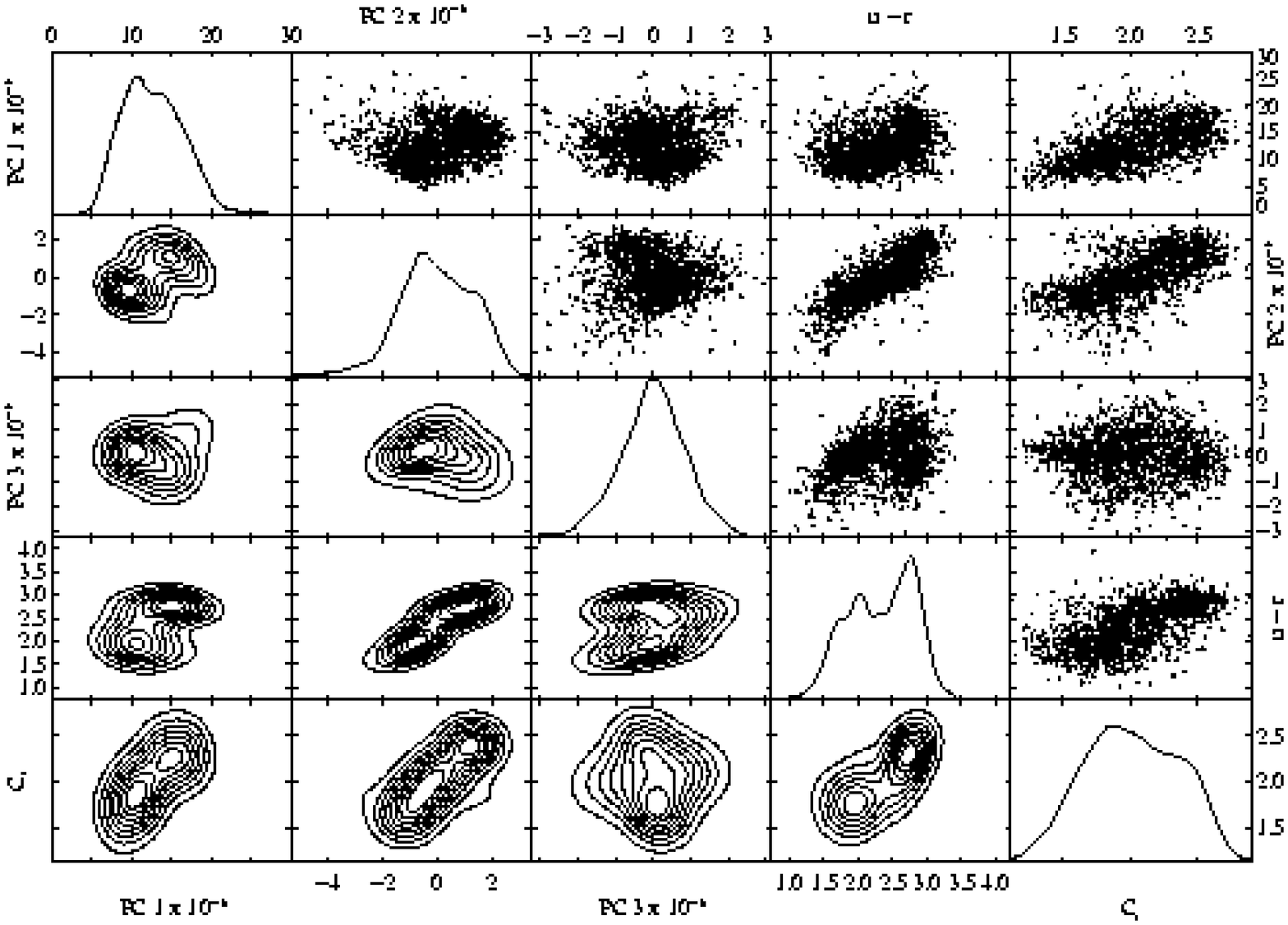}}}
\caption{Scatterplot matrix showing the distributions of $a_1, a_2, a_3, C_r,$ and $u-r$. The upper triangle plots show 2-dimensional scatterplots, the on-diagonal plots show the marginal probability densities, and the lower triangle plots show the 2-dimensional joint probability densities. \label{kl_vs_cur}}
\end{center}
\end{figure}

\clearpage
\begin{figure}
\begin{center}
\scalebox{1.2}{\rotatebox{90}{\plottwo{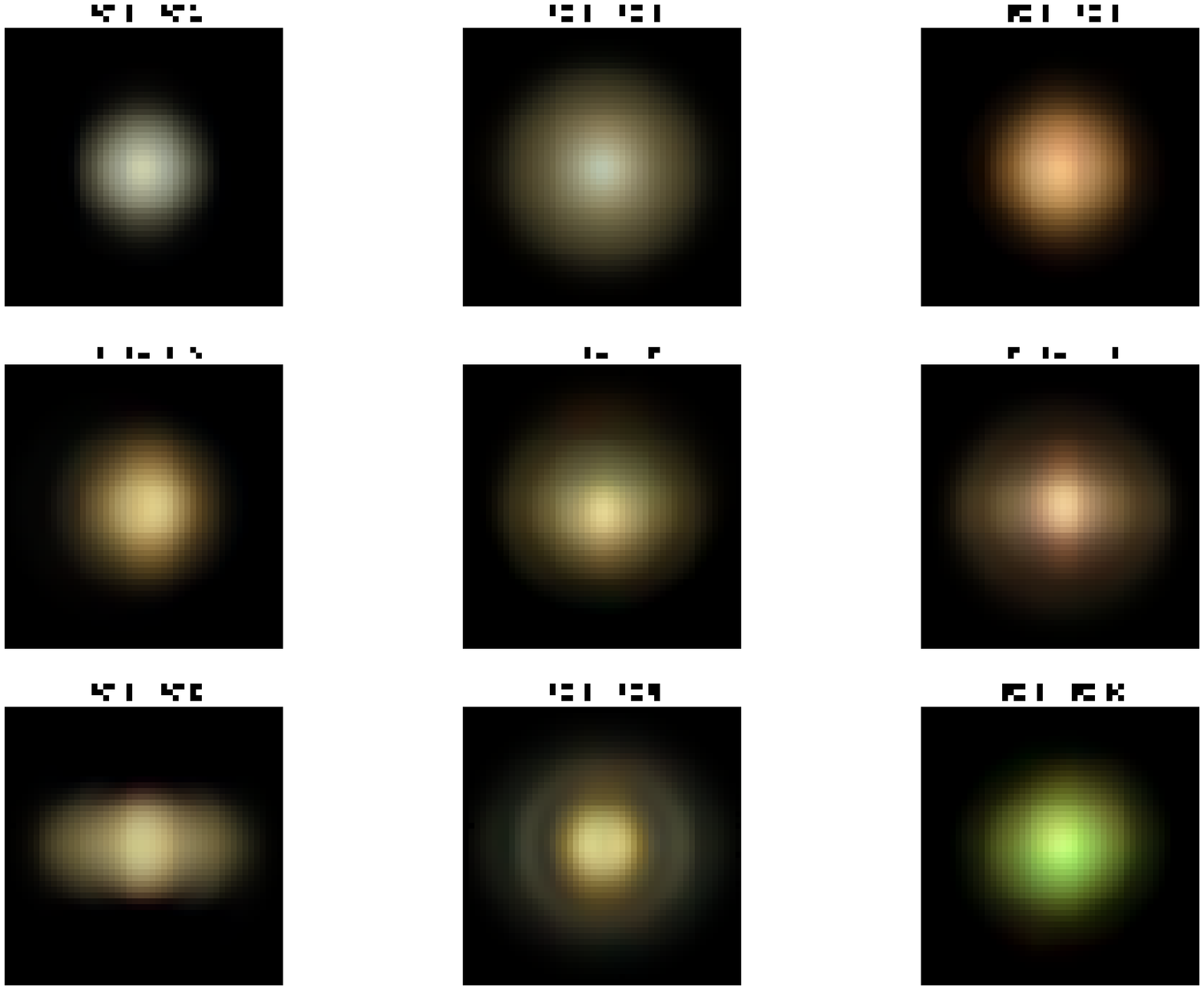}{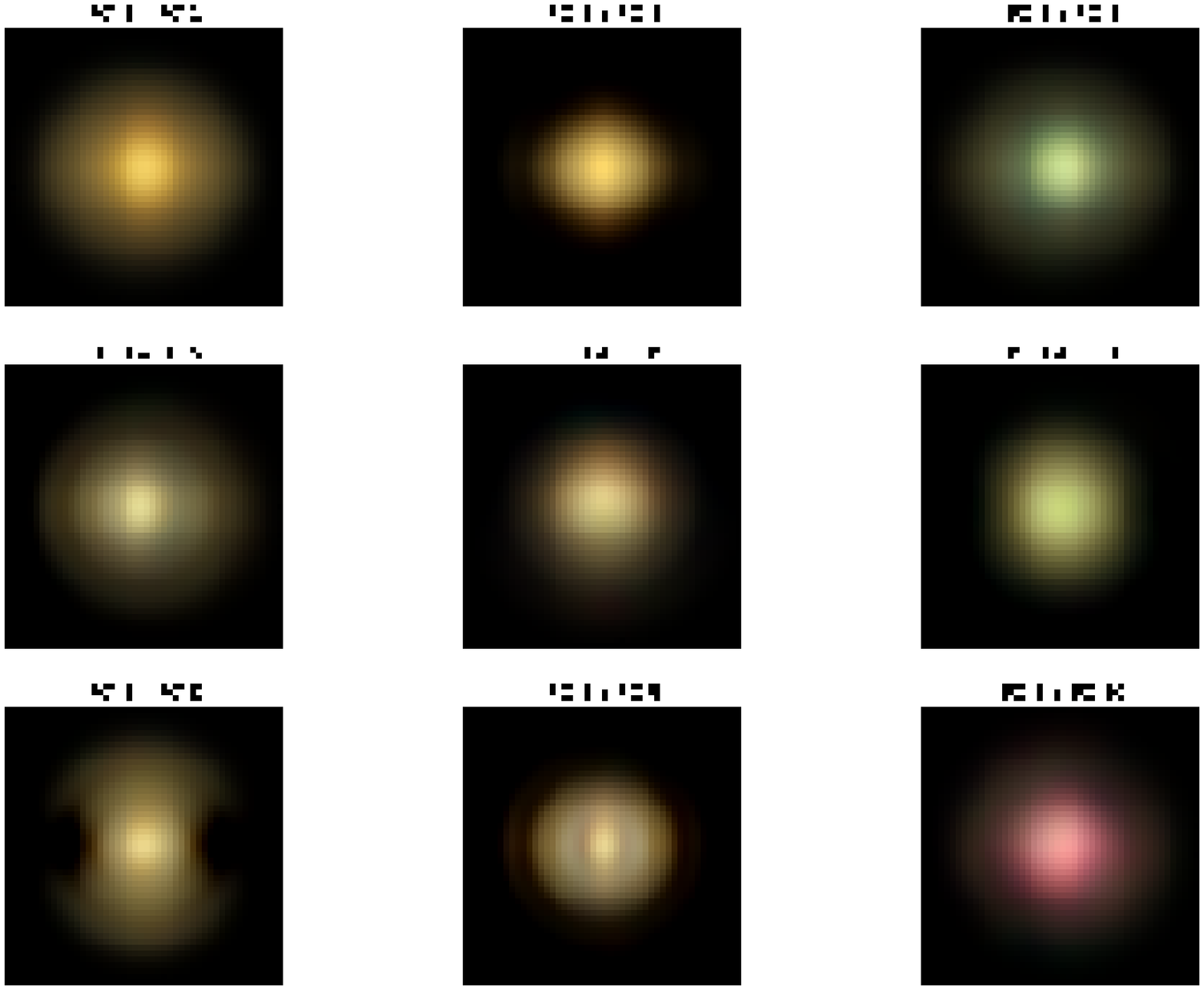}}}
\caption{RGB images showing the modifications of $v_1$ from the first ten $v_j, j>1$. Images are constructed from the $g, r,$ and $i$ band images using the asinh stretch. The absolute value of the coefficient, $|a_j|$, of each $v_j$ was chosen to be high, but not unrealistic, in order to emphasize the independent contributions of the $v_j$ to spectral morphology. \label{klmod}}
\end{center}
\end{figure}

\clearpage
\begin{figure}
\begin{center}
\scalebox{1.2}{\rotatebox{0}{\plottwo{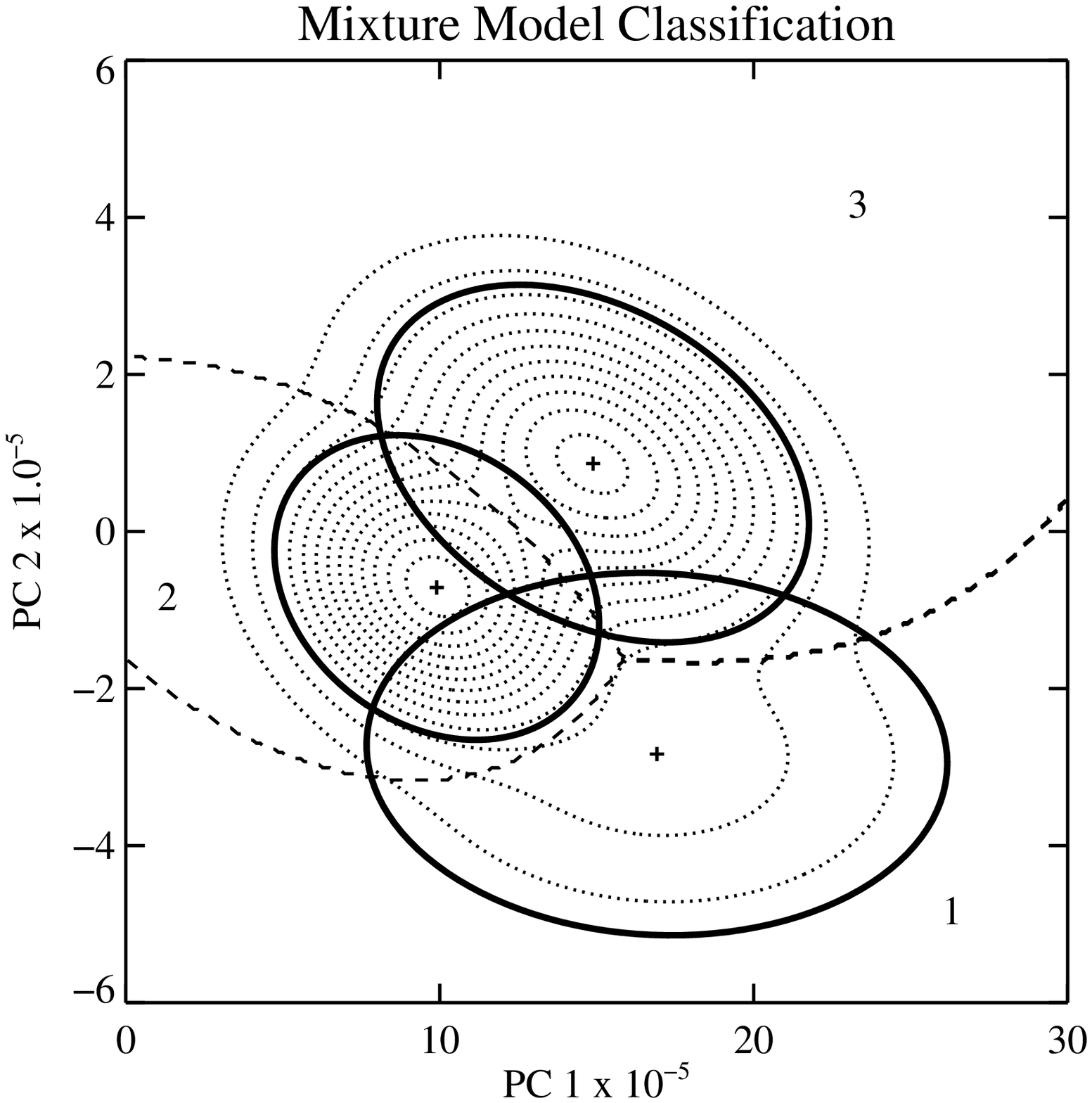}{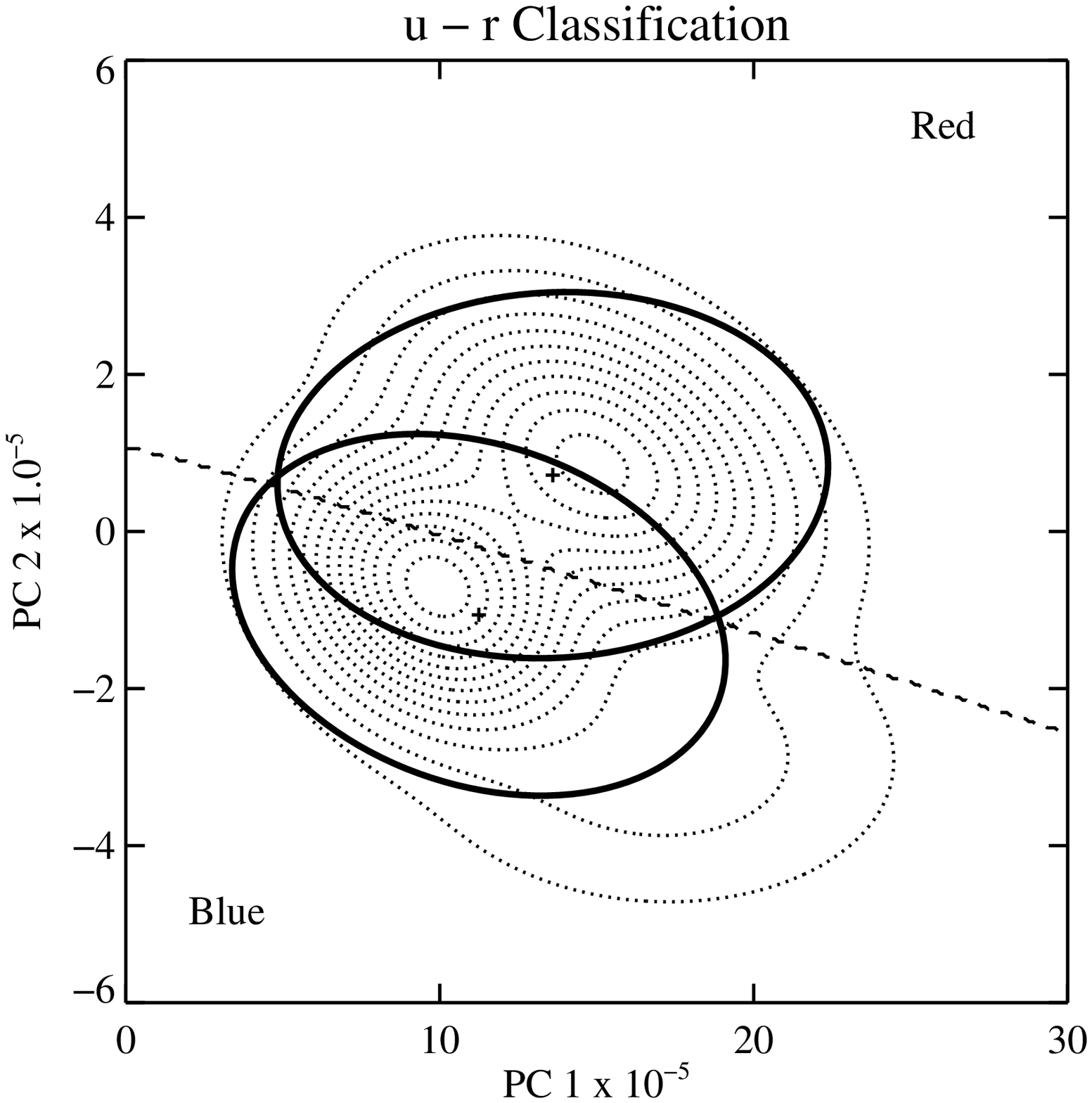}}}
\caption{The joint probability density of $a_1$ and $a_2$, as estimated by the mixture model fit. The dotted contours show the square root of the probability density. The square root is used in order to bring out the low probability density features, making the first mixture class more noticeable. The thick ellipses denote the regions of constant density that contain $95\%$ of the probability for each of the three mixture model classes (left) and the QDA fit to the red/blue $u-r$ classification (right). The dashed lines represent the decision boundaries for each of the mixture classes (left), and for the $u-r$ red and blue classes (right). The cross symbols mark the class centroids. As can be seen, the mixture classes are well separated in this plane. \label{dec_bounds}}
\end{center}
\end{figure}
\clearpage

\clearpage
\begin{figure}
\begin{center}
\scalebox{0.7}{\rotatebox{90}{\plotone{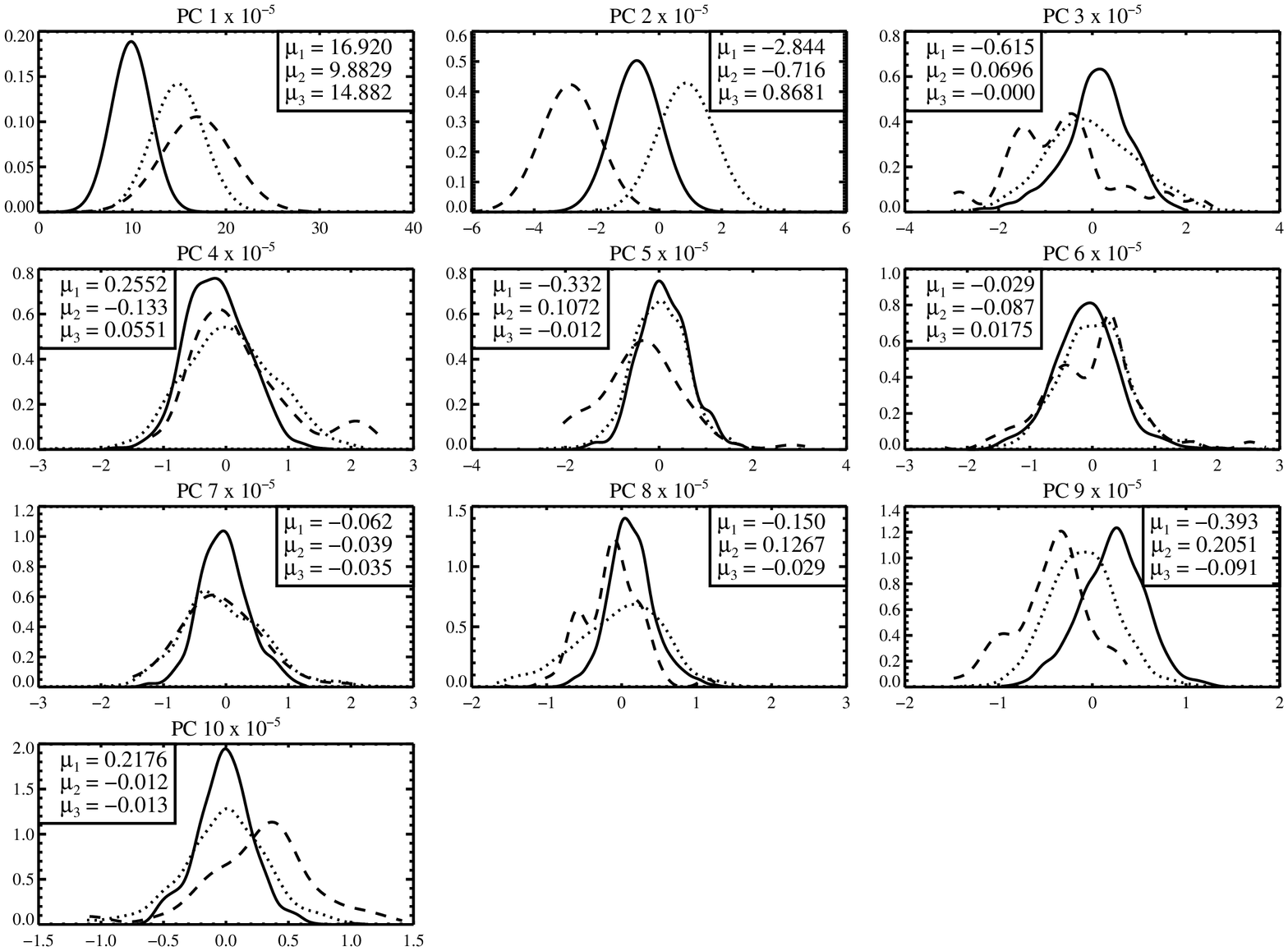}}}
\caption{Marginal probability densities of the mixture model classes in the first ten $a_j$. The densities have not been weighted by the respective class strengths, so the total probability densities will add up to three instead of one. We do this to show the individual marginal probability densities of each class, independent of the others. The densities for $a_1$ and $a_2$ are from the mixture model fit, the rest are estimated via kernel density estimation. The mean for the $k^{th}$ class is denoted $\mu_k$.  The dashed line represents $M_1$, the solid $M_2$, and the dotted $M_3$. \label{mixdens1d}}
\end{center}
\end{figure}

\clearpage
\begin{figure}
\begin{center}
\scalebox{0.7}{\rotatebox{90}{\plotone{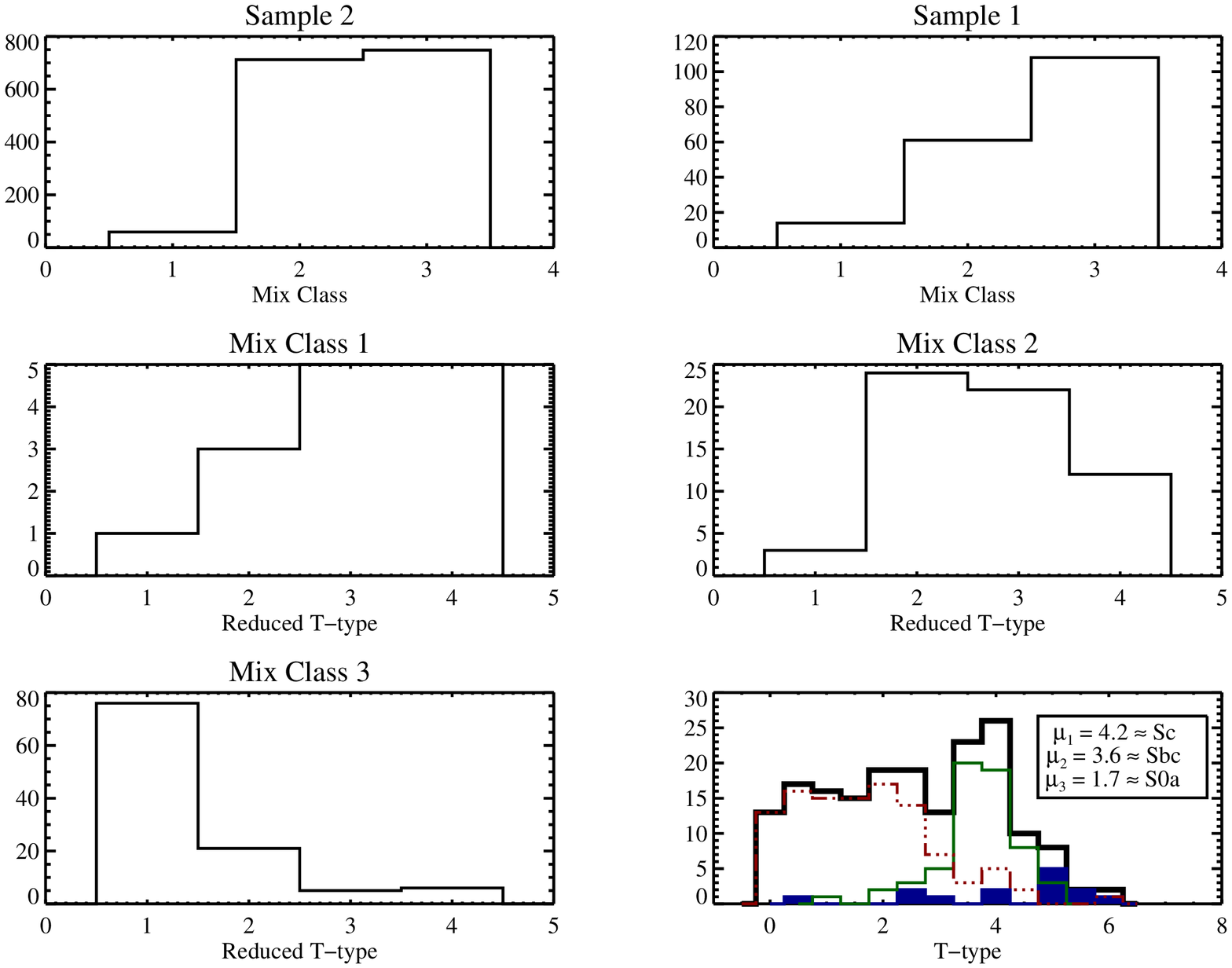}}}
\caption{Plots comparing the results of the mixture model classification to Hubble type.  The top two histograms show the distribution of mixture classes for the galaxies of Sample 2 and Sample 1, respectively.  The ``Reduced T-type'' seen in the following histograms is based on the Hubble classifications of Sample 1 and is coded as follows: 1=Early, 2=Middle, 3=Late, 4=Edge-on.  The T-type values of the bottom right histogram are also from the Sample 1 data, and are as follows: 0=E, 1=S0, 2=Sa, 3=Sb, 4=Sc, 5=Sdm, 6=Im. The blue filled histogram represents $M_1$ galaxies, the green thin solid line represents $M_2$ galaxies, and the red dashed-dot-dot-dot line represents $M_3$ galaxies.  The thick solid-lined histogram is for all Sample 1 galaxies combined. A clear division is seen between mixture classes 2 and 3 at a Hubble type of Sb; mixture class 1 appears to have an almost uniform distribution in Hubble type but is the most likely class occupied by Sdm and Im galaxies. \label{ttype_vs_mix}}
\end{center}
\end{figure}

\clearpage
\begin{figure}
\begin{center}
\includegraphics[angle=90, scale=0.45]{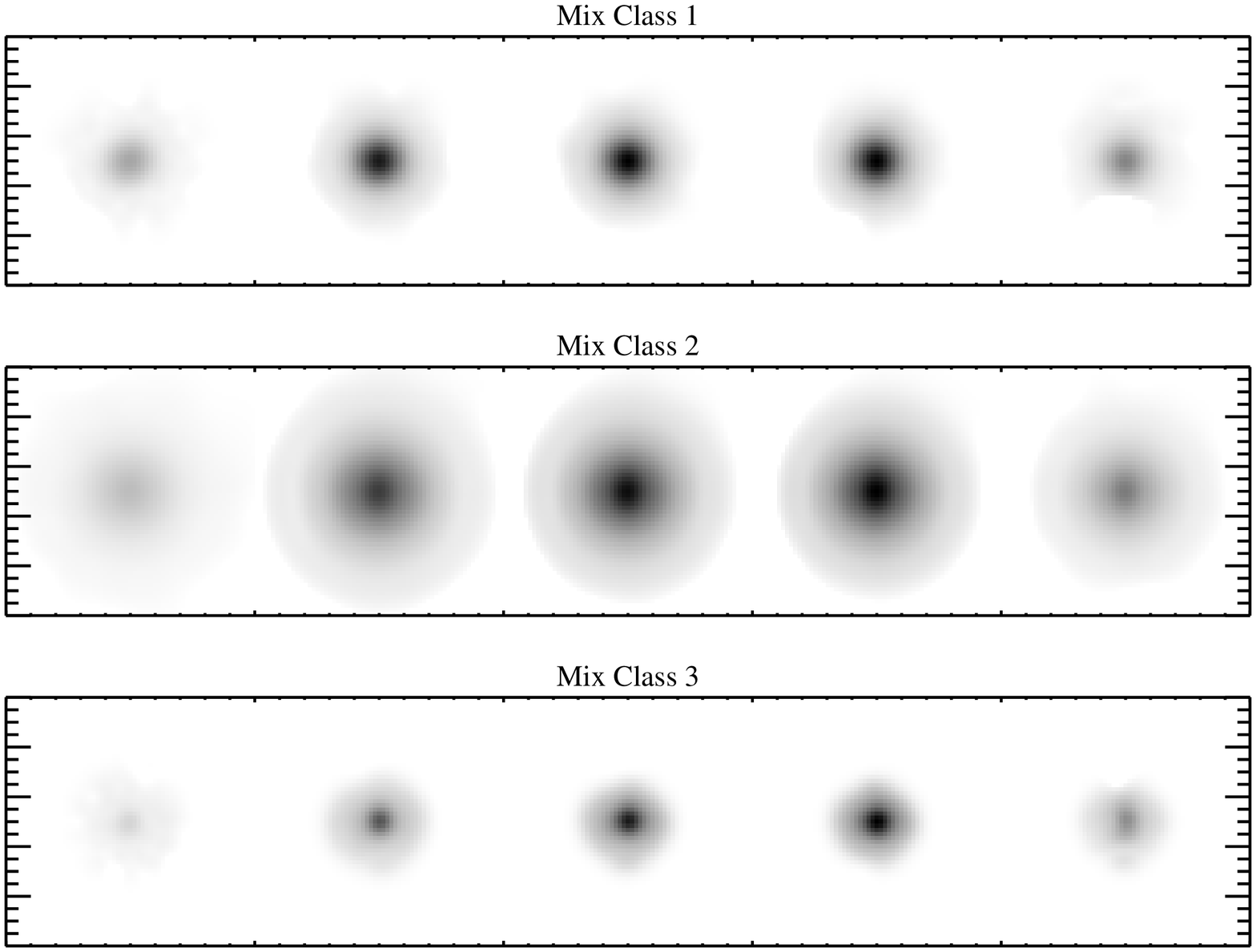}
\includegraphics[angle=0, scale=0.45]{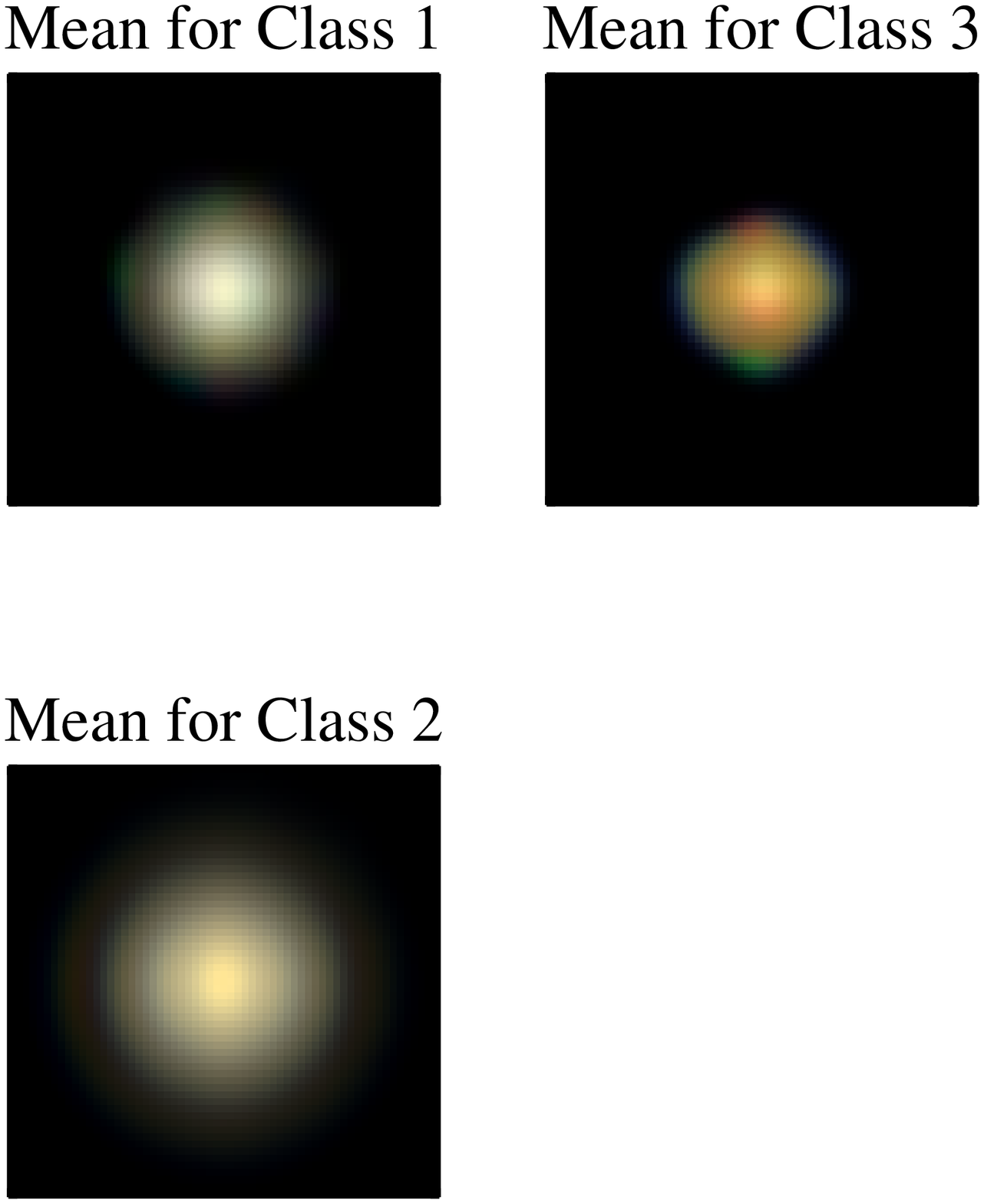}
\caption{The mean morphologies of the mixture model classes, reconstructed from the mean shapelet coefficients for each respective class.  As before, the bands of the images are (starting from the left) $u, g, r, i,$ and $z$, and the $g, r$, and $i$ bands are used for the RGB images. As usual, a square root stretch is used to show the black and white images, and the asinh stretch is used for the RGB images. \label{avgmix}}
\end{center}
\end{figure}

\clearpage
\begin{figure}
\begin{center}
\scalebox{1.2}{\rotatebox{90}{\plottwo{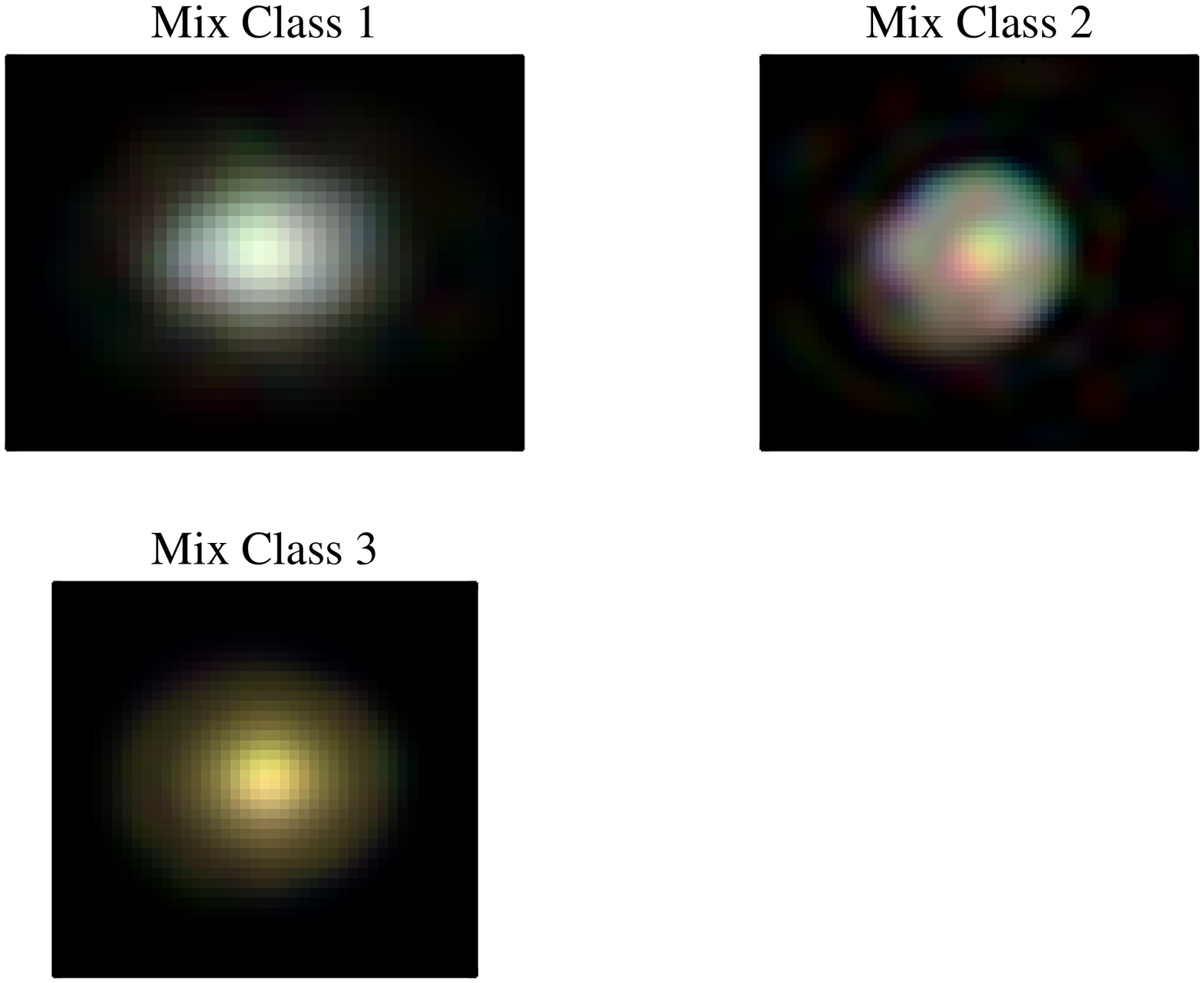}{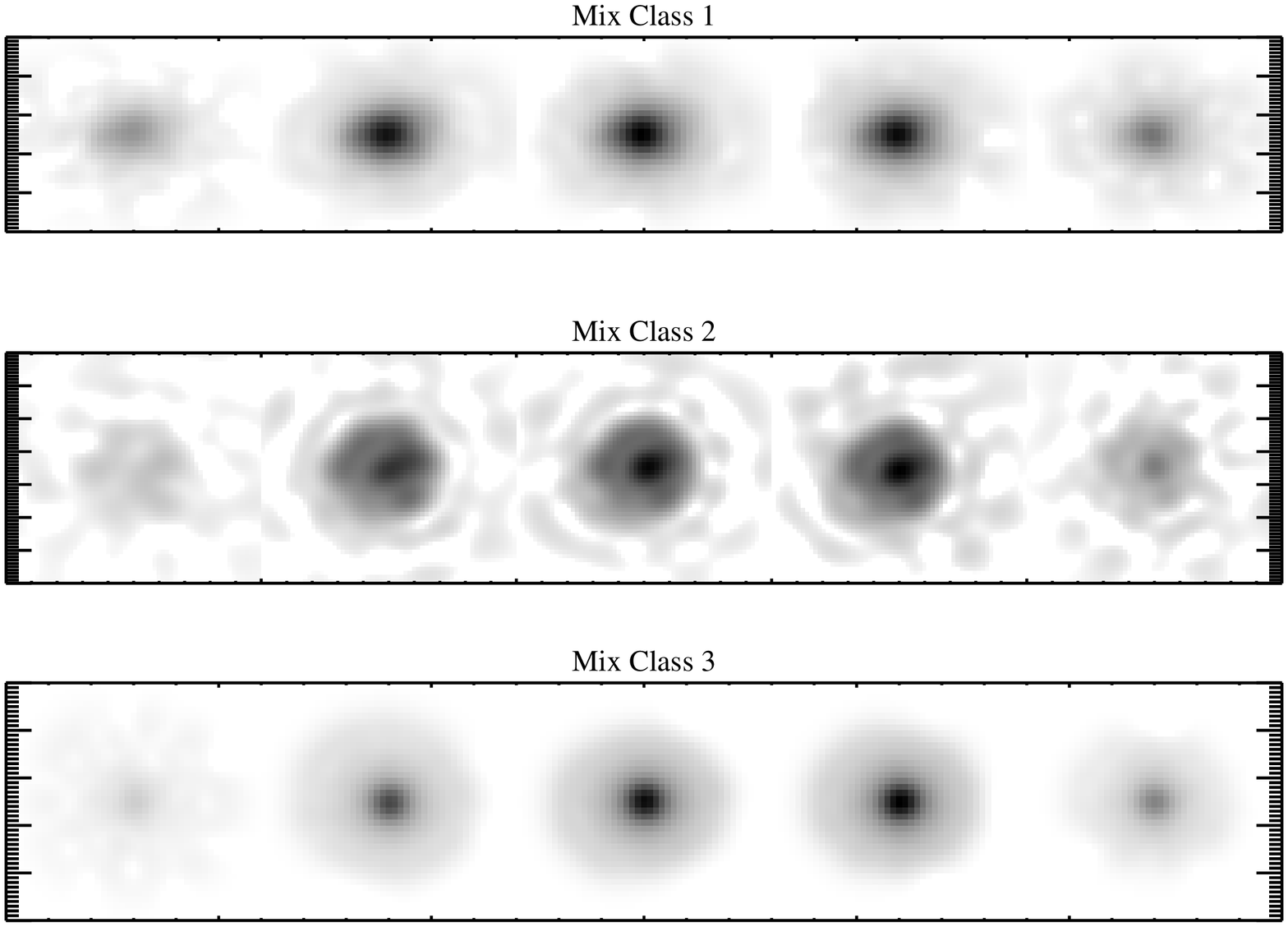}}}
\caption{Images of a galaxy from each $M_k$, constructed from their shapelet coefficients. The galaxies were chosen because they lie close to the mean vector for their respective classes, and so may be considered typical for their class. As usual, the bands of the RGB images are $g, r$, and $i$, and a square root stretch is used to display the black and white images. \label{mixgals}}
\end{center}
\end{figure}

\clearpage
\begin{figure}
\begin{center}
\scalebox{0.7}{\rotatebox{90}{\plotone{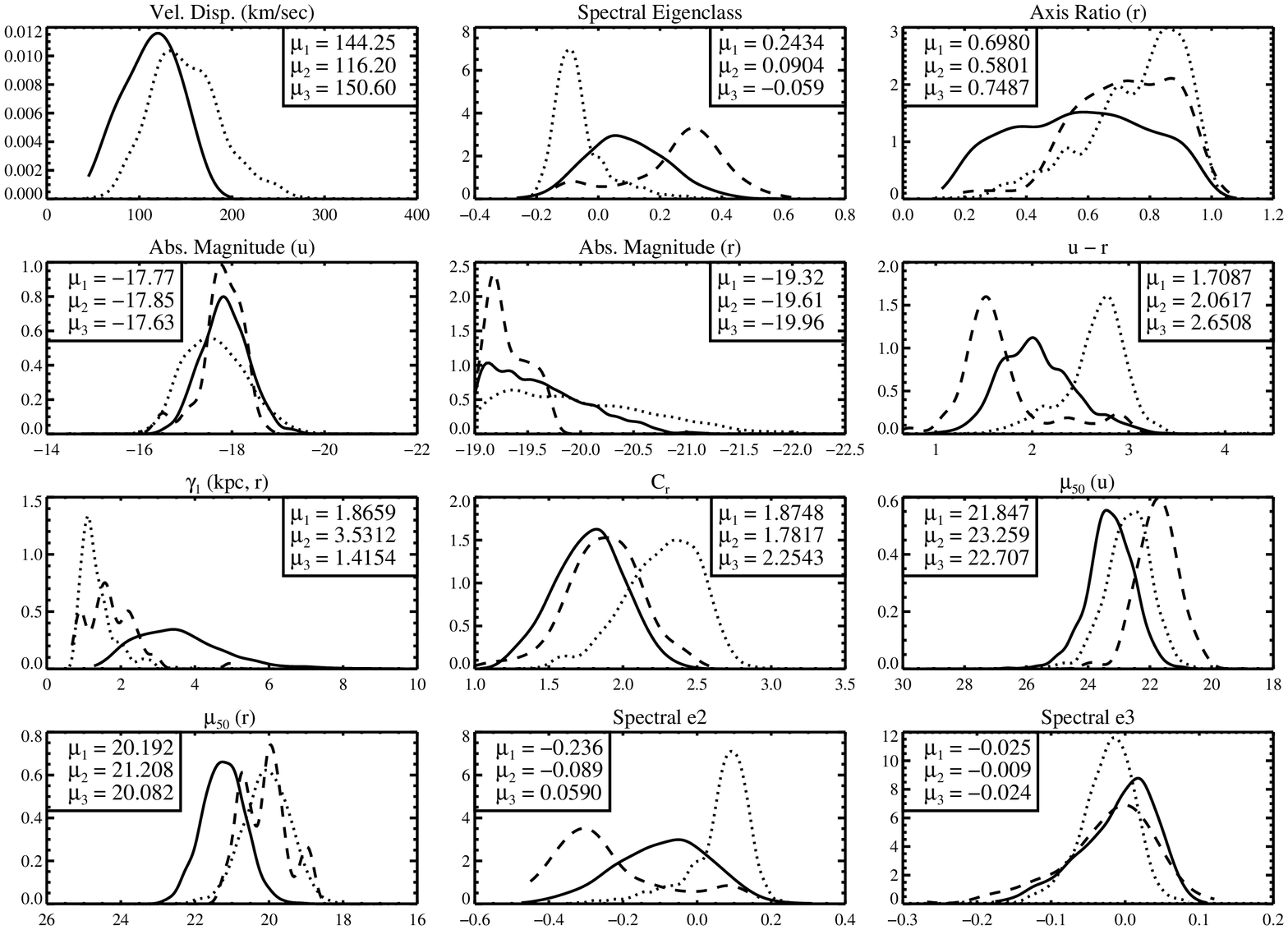}}}
\caption{The marginal probability densities of several physical parameters, shown separately for each mixture class. The class means are denoted by $\mu_k$. The dashed line represents $M_1$, the solid $M_2$, and the dotted $M_3$.  The $r$-band major axis shapelet scale before artificial redshifting and convolving (i.e., Eq.[\ref{eq07}]) in kpc is denoted by $\gamma_1$, the $r$-band concetration index is $C_r$, the half-light surface brightness is $\mu_{50}$, and the second and third spectral eigencoefficients are ``Spectral e2'' and ``Spectral e3.'' The SDSS spectral eigencoefficients are from the same analysis as the SDSS spectral eigenclass \citep{yip}. \label{phymix}}
\end{center}
\end{figure}

\clearpage
\begin{figure}
\begin{center}
\scalebox{0.7}{\rotatebox{90}{\plotone{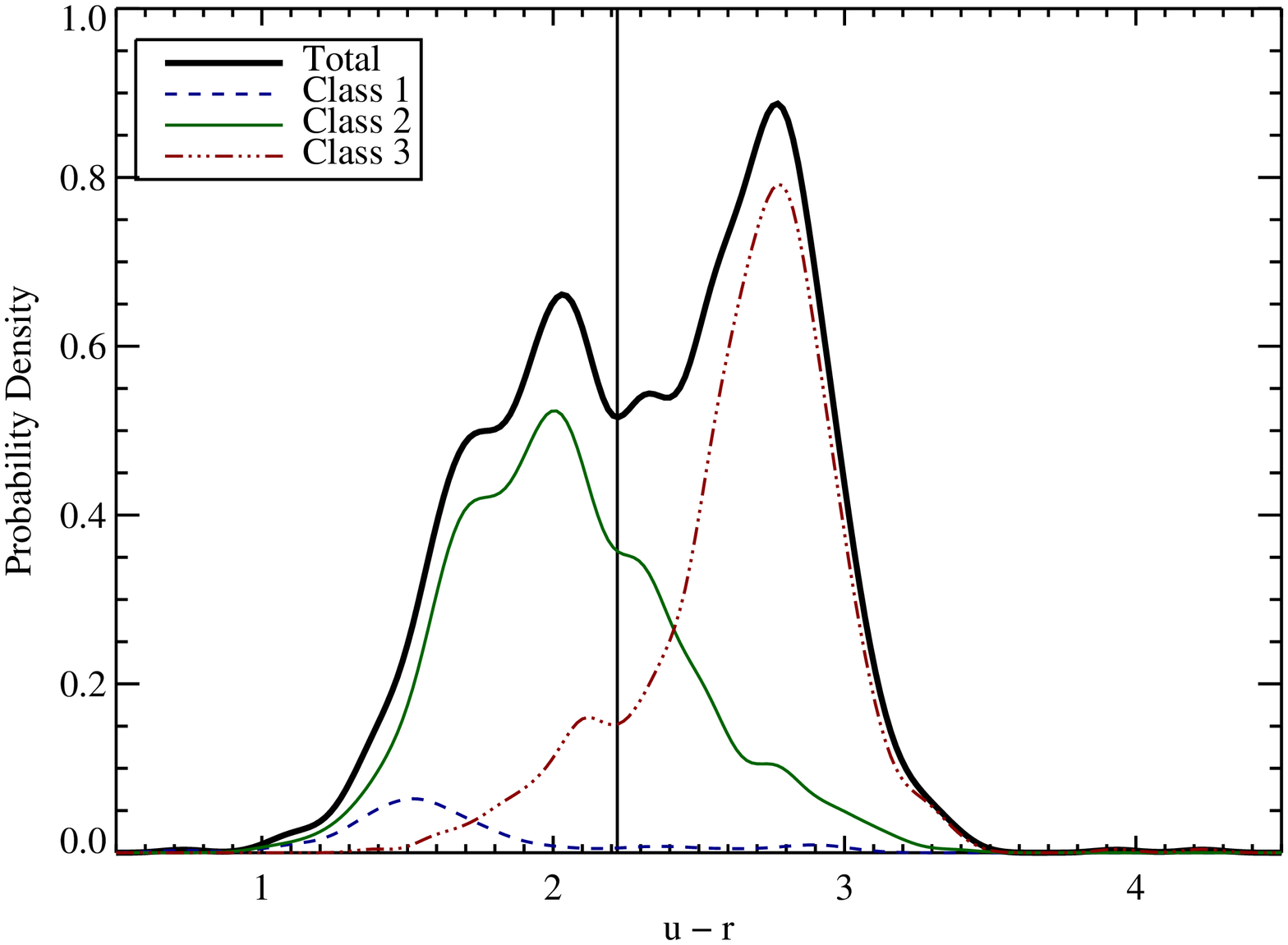}}}
\caption{Total $u-r$ probability density (thick black line), and the respective density components for $M_1$ (dashed blue), $M_2$ (thin green), and $M_3$ (dashed-dot-dot-dot red). The vertical line is the red/blue decision boundary at $u-r=2.22$.  \label{ur_vs_mix}}
\end{center}
\end{figure}

\clearpage
\begin{figure}
\begin{center}
\scalebox{1.2}{\rotatebox{90}{\plottwo{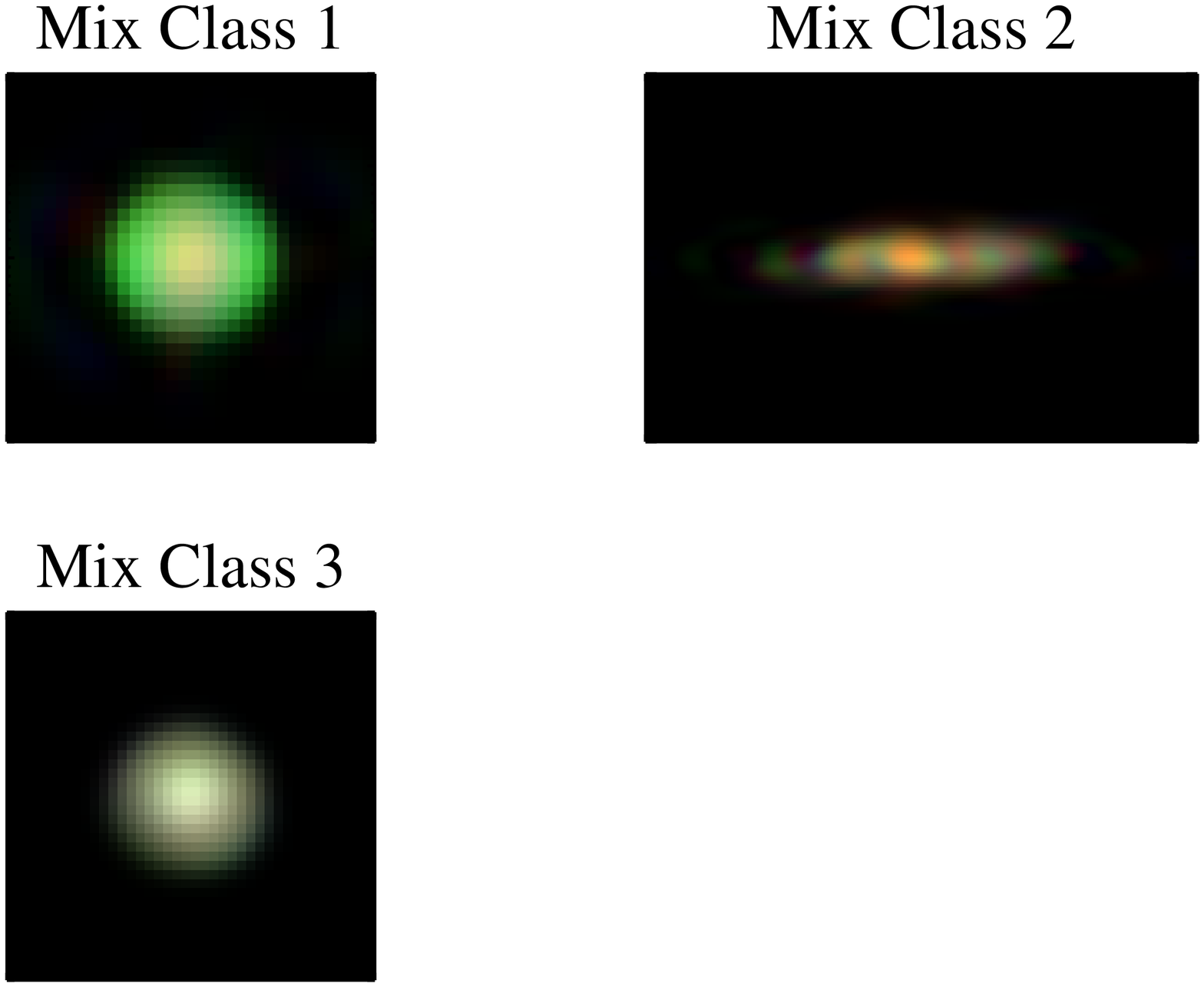}{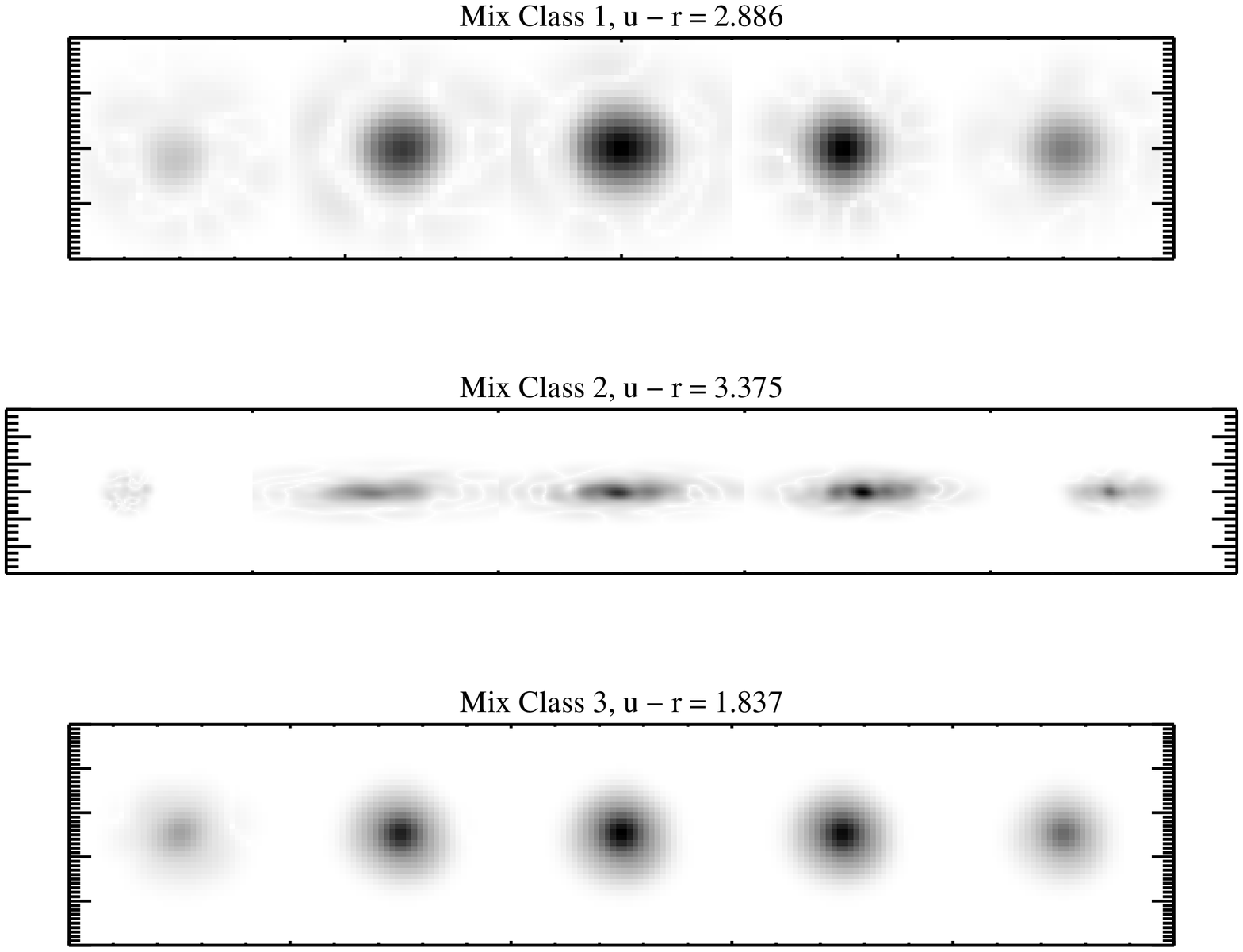}}}
\caption{Selected galaxies from each mixture class with a value of $u-r$ unexpected for that class. We show these images for the purpose of comparing with a simple cut on $u-r$ color. \label{odd_urgals}}
\end{center}
\end{figure}

\end{document}